\documentclass[letter,authoryear,12pt]{elsarticle}
\usepackage{amssymb,amsmath,stmaryrd,footmisc}
\usepackage[utf8]{inputenc}
\usepackage{esint}
\usepackage{placeins}
\usepackage{natbib}
\usepackage{nohyperref}
\usepackage{graphicx}
 \usepackage{setspace}
\usepackage{color}

\usepackage{etoolbox}
\makeatletter
\patchcmd{\ps@pprintTitle}
  {Preprint submitted to}
  {~}
  {}{}
\makeatother

\journal{~}
\begin{document}

\begin{frontmatter}
\title{Kinetics of surface growth with coupled diffusion and the emergence of a universal growth path}

\cortext[cor1]{Corresponding author: talco@mit.edu}

\author[label1]{Rami Abi-Akl}

\author[label1]{Rohan Abeyaratne}

\author[label1,label11]{Tal Cohen \corref{cor1}}

\address[label1]{Department of Mechanical Engineering, Massachusetts Institute of Technology, 77 Massachusetts Avenue Cambridge, Massachusetts, 02139, USA}
\address[label11]{Department of Civil \& Environmental Engineering, Massachusetts Institute of Technology, 77 Massachusetts Avenue Cambridge, Massachusetts, 02139, USA}

\begin{abstract}
Surface growth, by association or dissociation of material on the boundaries of a body, is ubiquitous  in both natural and engineering systems. It is the fundamental mechanism by which biological materials grow, starting from the level of a single cell, and is increasingly applied in engineering processes for fabrication and self-assembly. A significant complexity in describing the kinetics of such processes arises due to their inherent coupled interaction with the diffusing constituents that are needed to sustain the growth, and the influence of local stresses on the growth rates.\ Moreover, changes in concentration of  solvent within the bulk of the body, generated by diffusion, can affect volumetric changes, thus leading to an additional interacting growth mechanism.\ In this paper we present a general theoretical framework that captures these complexities to describe the kinetics of surface growth while accounting for coupled diffusion. Then, by combination of analytical and numerical tools, applied to a simple growth geometry, we show that the evolution of such growth processes rapidly tends towards a universal path that is independent of initial conditions. This path, on which surface growth mechanisms and diffusion act harmoniously, can be extended to analytically portray the evolution of a body from inception up to a treadmilling state, in which addition and removal of material are balanced.
 \end{abstract}


\end{frontmatter}

\section{Introduction}
\noindent Material growth is seldom governed by a single growth mechanism, but is rather a result of an interplay between several mechanisms that act in concert. Even in the simplest conceivable growth settings, such as 3D printing, formation of a body is often highly dependent upon chemo-mechanically coupled processes that occur within the material and can lead, for example, to shrinkage and residual stresses \citep{gibson, correa, szost,bauhofer2017direct,zurlo2017printing}. While 3D printing serves as an example  fabrication technique in which material is sequentially added on the boundaries of a body, other engineered processes may occur spontaneously, as in chemical deposition \citep{kaur, nair, mane, mattevi}, and growth of nanotube forests \citep{meyyappan,kim,louchev}. Such \textit{surface growth} mechanisms, are also  ubiquitous in nature and are  imperative for maintaining life. The cytoskeleton of animal cells grows by polymerization of actin gel on the inner side of the cell membrane\footnote{Actin gel is a fundamental building block of the cytoskeleton of mobile animal cells. It composes both the lamellipodium at the leading edge and the cell cortex and is the major engine of cell movement. Persistent polymerization of actin pushes against the cell membrane creating motion of the entire cell in a process which is controlled by the concentration of actin monomers.}, a surface growth process that is responsible  for cell motility \citep{mitchison,Mogilner96};  tissue level growth and remodeling of biological systems also involves material  deposition and removal on a surface \citep{ateshian2007} as for example, the renewal cycle of skin creating cell layers at the bottom surface of the hypodermis \citep{tepole2011,zollner2012}; seashells grow by deposition of material on an external surface \citep{Thompson,Skalak97,Moulton,Goriely}; and tree trunks grow by addition of layers underneath the bark \citep{archer2013}. In all of these examples, the grown body, although generated on a surface, can simultaneously evolve by chemo-mechanical processes occurring in the bulk that can, in-turn, influence the surface growth reaction. Eventually, the interplay between these  mechanisms  is what determines the evolution of the growing body and is thus the focus of the present study.

To consistently capture the complexities that arise  due to the existence of multiple reacting species in an open system,  the present approach  deviates from the more familiar \textit{kinematic growth theories}  \citep{Ambrosi2002, amar2005, Ambrosi2009, amar2010, Ambrosi2011, dervaux2011, Menzel,  Humphrey, Cyron} and attempts to represent the growth response by employing a thermodynamically consistent \textit{kinetic theory}. This is achieved by building on recent constitutive models for polymers that account for both large deformations and fluid diffusion \citep{Fried04,Hong,Duda,Chester,Loeffel}. Surface growth is then accounted for by permitting a chemical reaction between the two species (i.e. a solid matrix, and a permeating solvent), on the boundaries of the body. In general, we consider a solid matrix permeated by a solution that contains units that can associate to  (or dissociate from) the body. The solution can diffuse within the solid matrix and the resulting body can grow by a combination of two mechanisms, namely swelling and surface growth. Ultimately, we seek to understand the evolution of such a system, and to determine under what conditions the body evolves towards a steady-state in which addition and removal of solid material are balanced, also called treadmilling.

Although some available studies  model surface growth by considering a localized zone of  volumetric growth \citep{DiCarlo, Ciarletta,  Holland, Papastavrou}, thus facilitating the use of kinematic growth theories, in the present study we distinguish between  surface growth and volume growth. This dissimilarity has been emphasized by  \cite{Skalak97} and later illustrated by  \cite{Menzel} and is primarily due to the absence of a conventional natural global reference configuration in the case of surface growth, as manifested by the addition and removal of material points on the boundaries of the body and the resulting accumulation of residual stresses. The absence of a global natural reference configuration limits the application of usual continuum representations of the solid body,    and probably explains  the focus on kinematics in available studies of surface growth  \citep{Skalak82, Skalak97, Menzel, Moulton}. Recent studies \citep{TCA, sozio2017, ganghoffer2018} have  explored new ways to examine surface growth. Specifically, \cite{TCA} identified that for the particular surface growth scenario considered therein, the grown body possesses a natural global reference configuration if relaxed into a four-dimensional space. Building on this notion and extending it to an arbitrary surface growth geometry, in the next section we begin by defining the kinematic representation of the growth problem. Then in section \ref{Conservation}  we apply  the usual tools of continuum mechanics to formulate the governing equations of mass conservation and mechanical equilibrium  by alternating  between current and reference configurations to provide  a  measure for elastic deformation and stresses with respect to a stress-free state. In section \ref{Driving}, the definition of a natural reference configuration facilitates a thermodynamically consistent  derivation of configurational forces that drive association and dissociation. By examining those driving forces it will be shown that the growth reaction  is coupled to  particle diffusion, species concentrations, and internal stresses within the body. The theoretical framework is summarized in section \ref{Summary}, where a specific set of constitutive equations is provided to represent the growth of polymer gels. By applying these constitutive relations, the general theoretical framework is specialized to a simple growth geometry in  section \ref{1Dproblem}. It is shown that the evolution of such coupled growth processes rapidly tends towards a \textit{universal path} that is independent of initial conditions.\   Along this path the different growth mechanisms, i.e. swelling and surface growth, act harmoniously.

\section{Problem setting and kinematic representation}\label{Problem}
\noindent Consider a non-dilute solution of solvent units. Now, introduce into that solution an object of arbitrary shape that by chemical reaction  on its surface promotes  association  of solvent units to form a solid matrix.  That matrix can in-turn mix with the solvent and  swell to form an aggregate body composed of both a solid matrix and an impregnating solvent.  Since the binding reaction
can occur only on the growth surface, where it is energetically favorable, the previously formed layers are constantly pushed away from the growth surface as new layers are formed. This process may thus induce a build-up of internal stresses which can lead to subsequent dissociation. Additionally, a continuous supply of solvent units is required at the growth surface to sustain the
growth and to occupy the body region, implying that the growth reaction is intimately coupled with
the diffusion process.\footnote{A comprehensive nomenclature can be found in Appendix A.}

\begin{figure}[ht]
\begin{center}\includegraphics[width=0.6\textwidth]{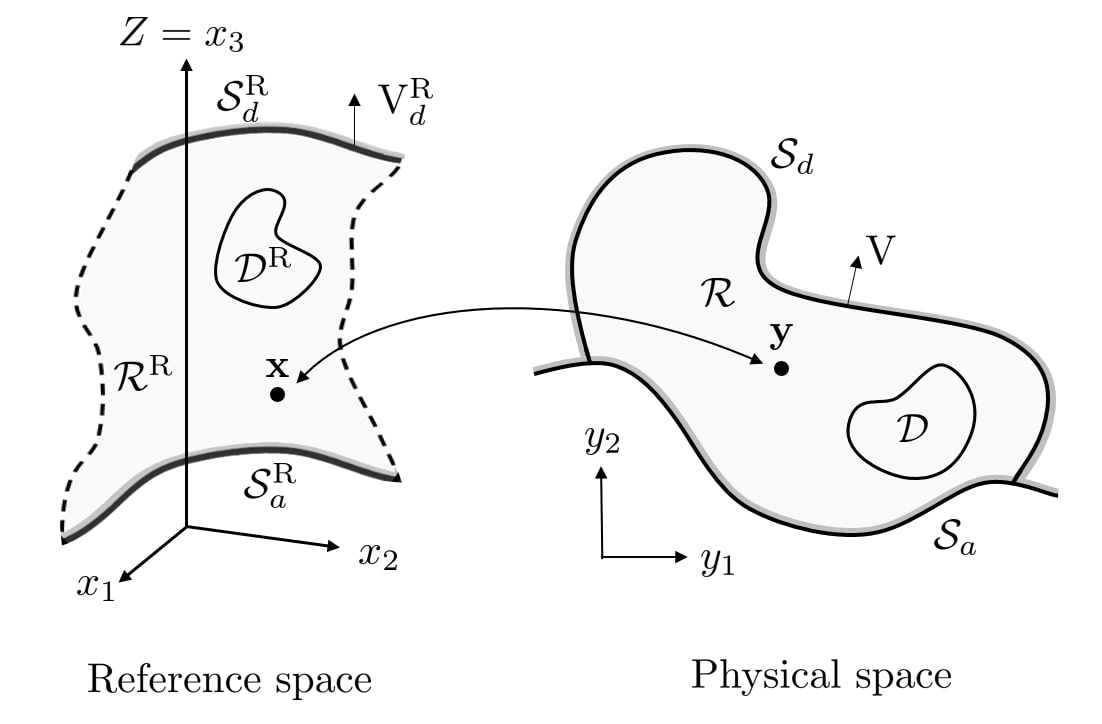} \end{center}
\caption{Schematic of the grown body in its reference and current $(\mathcal{R}^{\rm R}\mathcal{,R})$ configurations, respectively,  and the corresponding association $(\mathcal{S}^{\rm R}_a,\mathcal{S}_a)$ and dissociation  $(\mathcal{S}^{\rm R}_d,\mathcal{S}_d)$ surfaces. An arbitrary material subregion with its counterpart in the current frame    $(\mathcal{D}^{\rm R},\mathcal{D})$ is also shown as well as the boundary velocities in the reference and current frame,   $({\bf{V}}^R,{\bf{V}})$, respectively. For illustration purposes, the  grown body is represented here in a 2D physical space and its reference configuration is embedded in a 3D space.    }
\label{physical-reference}
\end{figure}

As the growth progresses, both the region  $\mathcal R(t)$  occupied by the body in the \textit{physical space} and the image region  $\mathcal R^{\rm R}(t)$
in the \textit{reference space} continuously evolve due to the reorganization of constituents (Fig. \ref{physical-reference}). Nevertheless,
though the region  $\mathcal R^{\rm R}(t)$
is evolving, each material point sustains its location in the reference space,
as long as it is attached to the body. As a convention, the superscripted `${\rm R}$' denotes values in the reference frame.

\bigskip

\smallskip

\noindent{\bf Physical space.}
In the physical space we represent the true, current configuration of the problem. A spatial point ${\bf y}$  is described by a set of orthonormal coordinates $(y_1,y_2,y_3)$. The region  $\mathcal R(t)$ occupied by the body in its  current configuration at time $t$ is referred to as the body manifold;  $\partial \mathcal R(t)$ denotes the surface boundary of the body manifold which can alternatively be written as $\partial \mathcal R(t)=\mathcal S_a(t)\cup \mathcal S_d(t)$, where $\mathcal S_a(t)$ and $\mathcal S_d(t)$ represent the association and dissociation surfaces respectively in the physical space at time $t$.
To relate between a volume element in its reference and current configurations it is useful to define an arbitrary material subregion   $\mathcal D(t)$ which is associated with a constant group of material points though solvent may diffuse in and out of it.\footnote{A smooth transition from association to dissociation can be facilitated by the definition of the kinetics of growth to avoid singularities at the intersection between $\mathcal S_a$ and $\mathcal S_d$. The specific discussion on singular solutions is beyond the scope of this work.}

\bigskip

\noindent{\bf Reference space.} In the reference space, the material manifold  $\mathcal R^{\rm R}(t) $  represents the solid matrix  in a stress-free and  dry state (i.e. without any solvent) and a material point ${\bf x}$ is described by a set of orthonormal coordinates $(x_i)$, with $i=1...n+1$, where $n+1$ denotes the dimension of the reference  space. Notice that the material manifold (including its boundary surface  $\partial \mathcal R^{\rm R}(t)$ and the
pre-image of    $\mathcal D(t)$  that is denoted by    $\mathcal D^{\rm{R}}(t)$) is  time dependent. This is a direct result of the association and dissociation on the boundaries.

In the above discussion we have taken the existence of
a reference configuration for granted. However, since the solid medium is continuously evolving,
with material particles being added and removed from the body, and with accumulation of residual
stresses, it is not a straightforward task to find an unambiguous global natural reference configuration.
Nevertheless, it is imperative to define the stress-free configuration at every point,
in order to evaluate the stress field in the growing material.\ To that end, a natural, geometrically
defined, reference configuration is essential. Though in the physical space such a configuration does
not necessarily exist, under certain conditions, it is possible to define a stress-free reference configuration in a
higher-dimensional space, as demonstrated by  \cite{TCA}, for the specific example of spherical
growth. Such a transformation can exist  if \textit{(i)} the growth is continuous
in time (i.e. addition of solid volume over time constitutes a continuous function), and \textit{(ii)} the
reference metric of the infinitesimally thin (2D) layer, formed on the growth surface at a given
time, can be isometrically embedded with finite bending in $\mathbb{R}^n$ with $n \geq3$. Hence, in the reference
space, the grown layers are hyper-surfaces and can be, intuitively, stacked on top of one another in
a chronological order along the $n+1$ dimension, to comprise the isometric embedding of the entire
body in $\mathbb{R}^{n+1}$. More specifically, if the embedding of the $2D$ grown layer in the reference space is defined by $\mathcal{S}_a^{\rm R}(t)\subset\mathbb{R}^n$, then the material manifold in the reference space is chosen as the Cartesian product \begin{equation}
   \mathcal{R}^{\rm R}(t)= \mathcal{S}_a^{\rm R}(t)\times{Z}(t),
\end{equation}where ${Z}(t)=\left[Z_a,Z_d\right]$ is the interval along the $n+1$ dimension, with $Z_a$ and $Z_d$ denoting the coordinate along the $n+1$ dimension of the association and dissociation surfaces respectively, and $\mathcal{R}^{\rm R}(t)\subset\mathbb{R}^{n+1}$. A wide group of problems can be limited to  \textit{simple growth surfaces} which we  define here as  growth surfaces that generate $2D$ layers that possess a stress-free reference configuration in $\mathbb{R}^3$. 

Having established that a global natural reference configuration of a body grown by the local mechanism
of surface growth can be defined, we proceed by employing such a transformation in a general
sense. Hence we define the mapping of a material point ${\bf x}$ in the reference space into a spatial point ${\bf y}$ in the physical space by the time dependent transformation
\begin{equation}
\label{position}
{\bf y}={\bf \hat y}\left({\bf x},t \right), \quad \quad {\bf x}\in {\mathcal R}^{\rm R}(t), \quad {\bf y}\in {\mathcal R}(t).
\end{equation}
The particle velocity and the deformation gradient tensor are respectively defined by
\begin{equation}\label{kinematics}
{\bf v}\left({\bf x}, t\right)=\frac{\partial \hat {\bf y}\left({\bf x}, t\right)}{\partial t}, \quad {\bf F}\left({\bf x}, t\right)=\frac{\partial \hat {\bf y}\left({\bf x}, t\right)}{\partial {\bf x}},
\end{equation} and
\begin{equation}
J({\bf x},t)=\det({\bf F}({\bf x},t)),\label{detF}
\end{equation} is the volume ratio\footnote{For simplicity, we will use the same notation, $J$, for both the volume ratio function expressed in terms of the reference frame coordinates, $J({\bf x},t)$, and the volume ratio function expressed in terms of the physical space coordinates, $J({\bf y},t)$.}.
In the presence of surface growth, distinction should be made between a point on the growth surface and the material point that happens to be at the same location at that instant. Let a point on the boundary of the body in the reference configuration be denoted by ${\bf x}_b(t)$, and its image in the physical configuration by ${\bf y}_b(t)={\bf \hat y}\left({\bf x}_b(t),t \right)$, the velocity of the boundary in the reference frame is then  ${\bf V}^{\rm R}={d{\bf x}_b}/{dt}$, and the true velocity of the boundary in the physical space is   ${\bf V}={d{\bf y}_b}/{dt}$.   By (\ref{kinematics})$^1$  we can now write the velocity of a solid unit that happens to be on the boundary  $({\bf x}_b(t))$ at the instant $t$ as
\begin{equation}\label{velocity}
{\bf v}={\bf V}-{\bf F}{\bf V}^{\rm  R}.
\end{equation}

\cite{Skalak82, Skalak97} provide a careful discussion on the kinematics of surface growth. In particular, they refer to the velocity with which the material points move away from the surface of growth as the growth velocity
\begin{equation} \label{growthvelocity}
{\bf V}_G={\bf v}-{\bf V}=-{\bf F}{\bf V}^{\rm R}.
\end{equation}
 In the absence of growth, ${\bf V}_{G}=0$ (or equivalently ${\bf V}^{\rm R}=0$), the boundary velocity and the material point velocity are equal. If a boundary is stationary in the physical space $({\bf V}=0)$, the material velocity and the velocity of the boundary in the reference frame are related by ${\bf v}=-{\bf F}{\bf V}^{\rm R}$.

In the general setting, the rate of addition of volume  on the boundary of the body, per unit area,  is thus

\begin{equation}\label{velocityrelation}
-{\bf V}_G \cdot {\bf n}\ dA_y=J{\bf V}^{\rm R}\cdot {\bf n}^{\rm R}\ dA_x,
\end{equation} where  $dA_y$ and $dA_x$ denote the  differential area elements in the physical and reference space, respectively,  that are related by the deformation gradient through
${\bf n}\  dA_y=J{\bf F}^{\rm -T}{\bf n}^{\rm R}\ dA_x
\label{surface}
$,
and with    ${\bf n}$ and   ${\bf n}^{\rm R}$ denoting the corresponding outward pointing normal vectors.
Hence, it is the normal component of the growth velocity that is associated with addition or removal of volume
  (${\bf V}_G \cdot {\bf n}<0$  corresponds to addition of volume) and from (\ref{velocityrelation}) it is apparent that addition or removal of volume in the reference configuration is directly related to the motion of the boundary   $({\bf V}^{\rm R}\cdot {\bf n}^{\rm R}>0$  corresponds to addition of volume).

As pointed out by \cite{Skalak82, Skalak97}, the growth velocity ${\bf V}_G$ need not be perpendicular to the growth
surface, and various growth forms may arise due to different directions of ${\bf V}_G$. In this study, we restrict attention to the case of normal growth in the reference configuration where ${\bf V}^{\rm R}={\rm V}^{\rm R}{\bf n}^{\rm R}$. This does not necessarily imply that in the physical space ${\bf V}_G$ is normal to the boundary $\partial \mathcal R (t)$; the in-plane component of the boundary velocity in the reference frame is attributed to the velocity of material points due to shear deformations that can be either externally imposed or chemically induced.

\section{Conservation laws and constitutive relations}\label{Conservation}
\noindent The considered system is composed of two species, a  solid matrix and a diffusing solvent. While on the boundaries of the body reactions of association or dissociation transform solvent units into solid units and vice versa, in  any sub-region within the bulk either species must be separately conserved. Following the frameworks described in \cite{Fried04}, \cite{Hong}, \cite{Duda} and \cite{Chester}  we require that each species is separately incompressible,  hence conservation of each species translates into conservation of volume.
 \smallskip

\noindent \textbf{Conservation of solvent.} Let $\phi=\phi({\bf y},t)$ denote the solvent volume fraction in the physical configuration, and ${\bf j}={\bf j}({\bf y},t)$  denote the true solvent flux (i.e. the volume of solvent  crossing a material unit area per unit time in the current configuration) the corresponding  referential quantities   $\phi^{\rm R}=\phi^{\rm R}({\bf x},t)$ and  ${\bf j}^{\rm R}={\bf j}^{\rm R}({\bf x},t)$ are related by the transformations

\begin{equation}\label{jjr}
\phi^{\rm R}={J}\phi\quad\text{and}\quad {\bf j}^{\rm R}=J{\bf F}^{-1}{\bf j.}
\end{equation}Then, within the body, conservation of solvent volume implies

\begin{equation}
\frac{\partial \phi}{\partial t}  + {\rm div} \, (\phi {\bf v}+{\bf j} )=0,
  \label{continuityphysical}
\end{equation}
in the current frame, or equivalently
\begin{equation}
\dot{\phi^{\rm R}} + {\rm Div}\left({\bf j}^{\rm R}\right) = 0,
\label{continuityreference}
\end{equation}in the reference frame.

\bigskip
\noindent {\bf Species balance.}
Since the solid units  and the solvent units are separately incompressible, changes in  the volume of a material region $\mathcal D(t)$ are solely due to the addition of solvent units. This can be written in integral form as
\begin{equation}
\frac{d}{dt} \int_{{\cal D}(t)} \phi \, dV_y = \int_{\partial {\cal D}(t)}  {\bf v} \cdot {\bf n} \ dA_y, \label{speciesbalance}
\end{equation}
and via usual integral transformations, can be rewritten in the local form as
\begin{equation}\label{localizedbalance}
\frac{\partial \phi}{\partial t} +
{\rm div} \, (\phi \,  {\bf v}) =  {\rm div}({\bf v}).
\end{equation}Combining the above relation with  the requirement of conservation of solvent units in \eqref{continuityphysical}  we arrive at the compact form of the species balance requirement
\begin{equation}
{\rm div} \, ({\bf v} +  {\bf j} ) =  0.
  \label{vjnur}
\end{equation}
Additionally, transforming (\ref{localizedbalance}) into the reference space,  yields
\begin{equation}\label{balanceReference}
\dot{\phi}^{\rm R} = \dot{J} ,
\end{equation}
and by integration we arrive  at a relation between the referential solvent volume fraction and the swelling ratio
in the form
\begin{equation}\label{phireference}
 \phi^{\rm R} =J-1,
\end{equation}
where, by definition and without loss of generality, we have set $J\equiv 1$ for the dry state ($ \phi^{\rm R}=0 $).
Combining \eqref{phireference} with the identity  \eqref{jjr}$^1$ we obtain the species balance relation
\begin{equation}\label{Jphi}
J=\frac {1}{1-\phi}  ,
\end{equation}which shows that the volume ratio of the solid matrix $J$ serves as a measure of the swelling and we thus refer to it as the swelling ratio. Substituting \eqref{Jphi} in \eqref{localizedbalance}, conservation of solvent volume can be expressed in terms of $J$ as
\begin{equation}
\label{dJdt}
\frac{1}{J^2}\frac{\partial J}{\partial t}={\rm div}\left(\frac{{\bf v}}{J}\right).
\end{equation}

Finally, equations \eqref{continuityphysical} and \eqref{vjnur}   are a sufficient and compact set of field equations to assure conservation of mass of both species {in any subregion within the bulk of the body, with  \eqref{Jphi}} relating between the volume ratio of the solid matrix and the solvent volume fraction. These equations must be complemented with considerations of mass conservation on the boundaries of the body.
\bigskip

\noindent{\bf Conservation of mass {on the boundaries of the body.}}
The association and dissociation reactions occurring on the boundaries of the body  locally invalidate the requirement of species conservation.  In other words, the flux of  solvent   flowing out\footnote{In the following,  quantities on the outer and inner sides of the boundary surface are indicated by a superposed $+$ and $-$ sign, respectively.} of the body through a unit area on its boundary   ${\bf j}^{+}\cdot{\bf n}\ dA_y$ need not be equal to   the flux of solvent flowing into the same unit area    ${\bf j}^{-}\cdot{\bf n}\ dA_y$. Nonetheless, assuming that the chemical reaction is isochoric\footnote{A non-isochoric reaction may be accounted for by introducing a local volume source.}, combined with the requirement that the species are separately incompressible, conservation of mass on the boundaries of the body translates to the requirement of volume conservation. Hence, returning to (\ref{vjnur}), in the presence of a singularity, conservation of volume across the surface reads
\begin{equation}
\left [\![ ({\bf v} +  {\bf j}\right)\cdot{\bf n}]\!] =0, \label{jin}
\end{equation}  where the square brackets denote the jump in the quantity across the boundary, such that  $[\![()]\!]=()^+-()^-$.
Since there is no material on the outer side of the boundary, ${\bf v}^+\equiv\bf 0$ by definition. To write an equivalent form of the above relation in the reference frame, we first notice that it can be rewritten as
\begin{equation}\label{jumpphysical}
[\![\-{\bf j}\cdot{\bf n}]\!]=[\![({\bf V}-{\bf v})\cdot {\bf n}]\!],
\end{equation}
then, using \eqref{growthvelocity}, \eqref{velocityrelation} and \eqref{jjr}$^2$,  conservation of volume across the surface reads
\begin{equation}\label{jumpreference}
[\![\-{\bf j}^{\rm R}\cdot{\bf n}^{\rm R}]\!]=\left(J-1\right){\bf V}^{\rm R}\cdot{\bf n}^{\rm R}.
\end{equation}Overall, the above equation (i.e. (\ref{jumpphysical}) or (\ref{jumpreference})) substitutes the requirements of species balance across the surface, by requiring that the local addition of solid volume is balanced by the local removal of solvent volume, to assure conservation of volume.

\bigskip

\noindent\textbf{Mechanical equilibrium.}  Let ${\bf T}={\bf T}({\bf y},t)$ and ${\bf S}={\bf S}({\bf x},t)$ denote the Cauchy and Piola stress tensors respectively, mechanical equilibrium requires that
\begin{equation}
{\rm div} \, {\bf T} + {\bf b} = {\bf 0}, \qquad {\bf T} = {\bf T}^{\rm T},  \label{Cauchy}
\end{equation}
\begin{equation}
{\rm Div} \, {\bf S} + {\bf b}^{\rm R} = {\bf 0}, \qquad {\bf SF}^{\rm T} =  {\bf FS}^{\rm T},  \label{Piola}
\end{equation}
where ${\bf b}={\bf b}({\bf y},t)$ and ${\bf b}^{\rm R}={\bf b}^{\rm R}({\bf x},t)$ denote the physical and referential body forces. Note that the Cauchy and Piola stress tensors are related by ${\bf S} =  J {\bf TF}^{\rm -T}$ and the physical and referential body forces are related by ${\bf b}^{\rm R} = J {\bf b}$.
\bigskip

\noindent{\bf Constitutive response.} The energetic state of a material unit can be defined by the elastic deformation of the matrix (represented by the deformation gradient tensor ${\bf F}$ ) and the solvent concentration (represented by the solvent volume fraction $\phi^{\rm R}$).  We thus represent the  free energy per unit referenial volume as
\begin{equation}\label{psi-definition}
\psi=\psi ({\bf F},\phi^{\rm R}).
\end{equation}Then, using the  Coleman-Noll methodology on the dissipation rate in a material subregion of the body, the constitutive relations can be derived. The  Cauchy and Piola stresses
read
\begin{equation}  \label{constitutiveCauchyPiola}
{\bf T}=J^{-1}\frac{\partial \psi}{\partial{\bf F}}{\bf F}^{\rm T}- p\,{\bf I}, \qquad {\bf S}=\frac{\partial \psi}{\partial{\bf F}}-p\,J{\bf F}^{-\rm T},
\end{equation} respectively, and the chemical potential per unit reference volume is

\begin{equation}\label{constitutivemu}
\mu= \frac{\partial \psi}{\partial {\phi^{\rm R}}}+p.
\end{equation}Here, since $\phi^{\rm R}$  is associated with the concentration of solvent, the \textit{hydrostatic pressure}\footnote{{This pressure can alternatively be referred to as an \textit{osmotic pressure}, since it does not necessarily vanish in absence of external loads and is a direct result of the species balance constrain that couples between the deformation of the solid matrix and the solvent intake}.  } $p$  arises as a reaction to the constrain \eqref{detF} and is constitutively indeterminate.

The dissipation in the bulk is  solely due to diffusion of solvent, hence the dissipation rate argument leads to the inequality ${\bf j}^{\rm R}\cdot{\rm Grad}\,\mu\leq0$. This inequality is satisfied by employing a kinetic law of the form
\begin{equation}\label{kineticlaw}
{\bf j}^{\rm R}=-{\bf M}({\bf F}, \phi^{\rm R})\,{\rm Grad}\,\mu,
\end{equation}where the mobility tensor ${\bf M}$ is positive semi-definite (i.e. $ \label{positiveM}
{\bf M}({\bf F}, \phi^{\rm R})\,{\bf g \cdot g} \geq 0 $
for all vectors ${\bf g}$).

Notice that according to the above kinetic relation, finite values of flux require  the chemical potential to be a continuous field. Specifically, this implies that on the boundaries of the body
\begin{equation}\label{mujump}
[\![\mu]\!] =0,
\end{equation}thus relating the chemical potential in the surrounding solvent to that in the body, at the interface.

\section{Driving force of growth}\label{Driving}
\noindent The constitutive equations  in the previous section describe the large deformation mechanics  of bodies composed of two species; a solid matrix and a diffusing solvent.   To tie between the bulk response and the growth kinetics that are localized at boundaries of the body, we return to the thermodynamic principles, and by considering the rate of dissipation associated with the growth reaction we seek to obtain the  driving force that governs growth. Since this force is responsible for changes in the material configuration, it can also be thought of as a configurational force \citep{gurtin2008configurational}.

We begin by writing the dissipation rate in the physical space by integration over the entire region
\begin{equation}\label{dissipationdef}
\begin{split}
\dot{\rm D}=\int_{\partial \mathcal R(t)}{{\bf T n}\cdot {\bf V}\ dA_y} +\int_{\mathcal R(t)}{{\bf b}\cdot {\bf v}\ dV_y}+\int_{\partial \mathcal R(t)}{\mu\left[-{\bf j}^{+}\cdot {\bf n}\ +({\bf V-v})\cdot {\bf n}\right]\ dA_y} \\-\frac{d}{dt}\int_{\mathcal R(t)}{\psi J\ dV_y},
\end{split}
\end{equation}
here the first two terms represent  the power invested by the external loads  acting on the boundary of  the body\footnote{Note that the velocity conjugate to the boundary traction is the boundary velocity $\bf V$ and not the particle velocity $\bf v$.}, and the body forces, respectively.   The third term characterizes the rate of energy flux into the system per unit time due to  solvent intake. The final term represents the rate of increase of the free energy of the system. Using the transformations \eqref{velocityrelation} and \eqref{jjr}$^2$ the dissipation rate can be  equivalently written
in the reference frame, as\begin{equation}\label{dissipation1}
\begin{split}
\dot{\rm D}=\int_{\partial \mathcal R^{\rm R}(t)}{{\bf S}{\bf n}^{\rm R}\cdot {\bf V}\ dA_x} +\int_{\mathcal R^{\rm R}(t)}{{\bf b}^{\rm R}\cdot {\bf v}\ dV_x}+\int_{\partial \mathcal R^{\rm R}(t)}{\mu\left(-{{\bf j}^{\rm R}}^{+}\cdot {\bf n}^{\rm R}\ +\ {\bf V}^{\rm R}\cdot {\bf n}^{\rm R}\right)\ dA_x} \\-\frac{d}{dt}\int_{\mathcal R^{\rm R}(t)}{\psi \ dV_x}.
\end{split}
\end{equation}
Now we proceed to reorganize \eqref{dissipation1}, to distinguish the dissipative mechanisms from the conservative terms.  As a first step, we rewrite the first two terms in  \eqref{dissipation1}   by  substituting relation \eqref{velocity}, then by employing the divergence theorem and inserting the  mechanical equilibrium relation \eqref{Piola} we arrive at the identity
\begin{equation}\label{OneAndTwo}
\int_{\partial \mathcal R^{\rm R}(t)}{{\bf S}{\bf n}^{\rm R}\cdot {\bf V}\ dA_x} +\int_{\mathcal R^{\rm R}(t)}{{\bf b}^{\rm R}\cdot {\bf v}\ dV_x}=
 \int_{\partial \mathcal R^{\rm R}(t)}{{\bf S}{\bf n}^{\rm R}\cdot {\bf F}{\bf V}^{\rm R}\ dA_x}+\int_{\mathcal R^{\rm R}(t)}{{\bf S}\cdot {\bf \dot F}\ dV_x}.
\end{equation}
Next,  the third term in  \eqref{dissipation1} can be rewritten with the aid of relation  \eqref{jumpreference} to obtain
\begin{equation}\label{term3}
\int_{\partial \mathcal R^{\rm R}(t)}{\mu\left(- {{\bf j}^{\rm R}}^{+}\cdot {\bf n}^{\rm R}\ +\ {\bf V}^{\rm R}\cdot {\bf n}^{\rm R}\right)\ dA_x}= \int_{\partial \mathcal R^{\rm R}(t)}{\mu\left(- {{\bf j}^{\rm R}}^{-}\cdot {\bf n}^{\rm R}\ +\ J{\bf V}^{\rm R}\cdot {\bf n}^{\rm R}\right)\ dA_x}.
\end{equation}
The first term in the above right-hand side integral can be further adapted by applying the divergence theorem and  using relation \eqref{continuityreference} to write
\begin{equation}\label{term3prime}
\int_{\partial \mathcal R^{\rm R}(t)}{\mu\left(-{{\bf j}^{\rm R}}^{-}\cdot {\bf n}^{\rm R}\right)\ dA_x}= \int_{\mathcal R^{\rm R}(t)}{\left(-{{\bf j}^{\rm R}}^{-}\cdot {\rm Grad}\ {\mu }\ +\mu \dot \phi^{\rm R} \right)dV_x}.
\end{equation}
There remains the final term that can be reexpressed using the transport theorem as
\begin{equation}\label{term4}
\frac{d}{dt}\int_{\mathcal R^{\rm R}(t)}{\psi \ dV_x}=\int_{\mathcal R^{\rm R}(t)}{\dot \psi \ dV_x}-\int_{\partial \mathcal R^{\rm R}(t)}{\Delta \psi {\bf V}^{\rm R}\cdot {\bf n}^{\rm R}\ dA_x},
\end{equation}
where
\begin{equation}\label{latentenergyDef}
\Delta \psi = \psi^{+}-\psi^{-},
\end{equation}
represents the jump in potential energy density  across the moving boundary. This difference will be referred to as the latent energy of growth. It arises due to the mechanism of local energy transduction associated with the phase transformation (from solvent phase into the solid phase or vice versa).

Recalling that the free energy density  $\psi = \psi ({\bf F}, \phi^{\rm R})$ is taken to be a function of the deformation gradient tensor ${\bf F}$ and the referential volume fraction  $\phi^{\rm R}$, we can write
\begin{equation}\label{psidot}
\dot \psi = \frac{\partial \psi}{\partial {\bf F}}\cdot {\bf \dot F}+\frac{\partial \psi}{\partial \phi^{\rm R}}{\dot \phi^{\rm R}}={\bf S \cdot \dot F}+\mu \dot \phi^{\rm R}.
\end{equation}

Now, by substituting relations \eqref{OneAndTwo}-\eqref{term4} and \eqref{psidot} in \eqref{dissipation1} and reorganizing, the dissipation rate reads
\begin{equation}\label{totaldissipation}
{\rm \dot D}=\int_{\partial \mathcal R^{\rm R}(t)}{\left({\bf F}^{\rm T}{\bf S}\ + \Delta \psi \ {\bf I}+\mu {J}\ {\bf I} \right){\bf n}^{\rm R}\cdot {\bf V}^{\rm R} \ dA_x}-\int_{\mathcal R^{\rm R}(t)}{{\bf j}^{\rm R}\cdot {\rm Grad}\ {\mu } \ dV_x}.
\end{equation}
Here the two integrals characterize the two different mechanisms of energy dissipation: the surface integral is associated with dissipation on the boundary and the volume integral with dissipation in the bulk due to the diffusion of solvent through the solid matrix.

The constitutive relation in \eqref{kineticlaw} assures us that the dissipation rate associated with diffusion is non-negative, hence the remaining inequality that must be satisfied to guarantee a non-negative dissipation rate, is

\begin{equation}\label{Dgrowth1}
\int_{\partial \mathcal R^{\rm R}(t)}{\left({\bf F}^{\rm T}{\bf S}\ + \Delta \psi \ {\bf I}+\mu {J}\ {\bf I} \right){\bf n}^{\rm R}\cdot {\bf V}^{\rm R} \ dA_x}\geq0.
\end{equation}
As explained in Section \ref{Problem}, in the present study we restrict attention to the case of normal growth in the reference configuration such that, ${\bf V}^{\rm R}={\rm V}^{\rm R}{\bf n}^{\rm R}$. {Recall that this restriction  does not necessarily imply normal growth in the physical configuration, but excludes growth induced  shear deformations at the surface.} Hence, the dissipation inequality \eqref{Dgrowth1} can be rewritten as
\begin{equation}\label{Dgrowth2}
\int_{\partial \mathcal R^{\rm R}(t)}{\left({\bf S}{\bf n}^{\rm R}\cdot {\bf F}{\bf n}^{\rm R}\ + \Delta \psi +\mu {J}\right){\rm V}^{\rm R} \ dA_x}\geq0.
\end{equation}

Finally, from the above relation we identify the thermodynamic conjugate to the growth rate $({\rm V}^{\rm R}{\bf =V}^{\rm R}\cdot {\bf n}^{\rm R})$ which is  the  \textit{driving  force of growth} that acts along the growth direction (per unit reference area)
\begin{equation}\label{drivingforce}
f={\bf S}{\bf n}^{\rm R}\cdot {\bf F}{\bf n}^{\rm R}\ + \Delta \psi +\mu {J}.
\end{equation}
The dissipation inequality \eqref{Dgrowth2} thus reduces to the requirement
\begin{equation}\label{fVpositive}
 f{\rm V}^{\rm R}\ge 0 \quad {\rm on} \quad {\partial \mathcal R^{\rm R}(t) }.
\end{equation}
Recalling that the boundary velocity in the reference frame is such that ${\rm V}^{\rm R} > 0$ for association, and ${\rm V}^{\rm R}<0$ for dissociation, we thus infer from inequality \eqref{fVpositive} that the driving force is positive  $f > 0$  for association, and negative $f<0$ for dissociation.

Examining the driving force relation (\ref{drivingforce})  we identify three terms. The first term is associated with   mechanical forces acting on the boundary.   The second term is the latent energy of the growth reaction  and the third term is the  influx of chemical energy. These three effects act together to define the kinetics of surface growth. Note that a similar result for the driving force was obtained by \cite{TCA}. Therein, the coupling between growth and diffusion was accounted for through an external field and not coupled to the mechanical response, hence the difference in the driving force terms.

\bigskip

\noindent\textbf{Kinetic law of growth.}
In light of  inequality in \eqref{fVpositive}  a specific form of the kinetic law that governs growth can be chosen as  \begin{equation}\label{fVRcondition}
{\rm V}^{\rm R}=\mathcal{G}\left(f \right),
\end{equation}where the growth function  $\mathcal{G}\left(f \right)$ obeys the inequality    $f\mathcal{G}\left(f \right)\ge 0 $ for all values  of $f$.

\section{Summary of formulation and specific constitutive equations}\label{Summary}
\noindent In the previous sections we have developed a complete formulation of the mechanics of a body composed of two species (i.e. solid and solvent) that can grow by combination of two mechanisms (i.e. swelling and surface growth). In the physical space, the body is described in spatial coordinates $\bf y$, and its  boundaries  ${\bf y}_b$ have velocity $\bf V$. Equivalently, in the reference frame, the body can be described using coordinates $\bf x$ and ${\bf x}_b$ and the boundary velocity ${\bf V}^{\rm R}$. Fields $\bf v$, $\bf F$ and $J$ are key to relate between the two configurations and to thus define the state of the body. Mass conservation in the physical space is governed by the continuity equation \eqref{continuityphysical}, the incompressibility argument \eqref{vjnur} and the species balance \eqref{Jphi}, to link the fields $\bf v$ and $J$ to  the solvent volume fraction $\phi$ and the solvent flux $\bf j$. Or equivalently, in the reference space, equations \eqref{continuityreference} and \eqref{phireference} relate  $J$ to $\phi^{\rm R}$ and ${\bf j}^{\rm R}$.  Mechanical equilibrium is accounted for in the current configuration through  \eqref{Cauchy}, and in the reference configuration through   \eqref{Piola}. The stress tensors  $\bf T$, $\bf S$ and chemical potential $\mu$ obey the constitutive relations \eqref{constitutiveCauchyPiola} and  \eqref{constitutivemu}, respectively and depend on  a specific form of the free energy $\psi({\bf F}, \phi^{\rm R})$. The diffusion law \eqref{kineticlaw} relates the referential flux  ${\bf j}^{\rm R}$ to the gradient of chemical potential through a mobility tensor ${\bf M}({\bf F}, \phi^{\rm R})$ and the physical flux  $\bf j$ can be derived through the equivalence relation \eqref{jjr}$^2$.

On the boundaries of the body, we shall account for the volume conservation across the surface \eqref{jin},  the continuity of  chemical potential \eqref{mujump}, and the applied loads or physical constraints. The density of material  growing on a given growth surface is a property of this surface (it can be influenced by surface roughness and chemical binding properties). This surface density predetermines the tangential deformation of the solid matrix on the association surface and provides an additional boundary condition. Finally,  the kinetic law of growth $\mathcal G(f)$ must be defined to determine the growth kinetics and is related to the  driving force  \eqref{drivingforce} and the growth velocity \eqref{fVRcondition}.

In summary, for a given material characterized by a constitutive response defined by $\psi$ and  ${\bf M} $,   the kinetic relation  $\mathcal G$ is the only additional constitutive relation needed to specialize the  growth kinetics to a specific problem. Before we proceed to apply our model to a boundary-value problem we thus choose a specific set of
constitutive relations. We consider a setting in which the grown material is a polymer gel that is composed of a polymer network and an impregnating solvent that contains monomers. Polymer gels that can grow and swell are of particular interest due to their ubiquity in natural systems, additionally   they have been shown to associate and dissociate  on their boundaries in artificial settings \citep{Noireaux,bauer2017new} thus providing a model system for the investigation of chemo-mechanically coupled growth.

\parindent=2em

\
\\
\textbf{Constitutive response in the bulk}. To account for both finite stretching of the polymer network and for species migration, we employ a constitutive framework for the bulk response based on the models suggested by \cite{Hong}, \cite{Chester}, and \cite{Duda} with some minor modifications in their representation.

We construct the Helmholtz free energy following the {\cite{Flory43} approach} by  accounting separately for the contribution of each species to write
\begin{equation}\label{psi}
\psi ({\bf F},\phi^{\rm R})= \psi_{{e}}({\bf F})+ \psi_{{s}}(\phi^{\rm R}),
\end{equation}where  $\psi_{{e}}({\bf F})$ represents the free energy of the solid matrix due to elastic deformation, and $\psi_{{s}}(\phi^{\rm R})$ represents the free energy of the solvent and depends on the fraction of solvent within a volume  unit given by $J$. The former is taken as  \citep{Treloar}
\begin{equation}\label{psie}
\psi_e({\bf F})=\frac{G}{2}\left[{|{\bf F}|}^2-3 -2 \ln \left(\det {\bf F} \right)\right], \quad G=NkT,
\end{equation}
where $G$ is the elastic shear modulus, $N$ represents the number of solid chains per unit volume of the body, $k$ is the Boltzmann's constant, and $T$ is the  temperature.

 Assuming that the unmixed solvent  in the surrounding region is in chemical equilibrium, it has a constant chemical potential $\mu_0$ and   due to incompressibility, the  free energy of a unit volume of unmixed solvent is
$\psi_0=\mu_0 $. 
Once the solvent penetrates into the solid matrix it changes its free energy (per unit volume) by both change in relative volume and by mixing. The free energy of mixing is assumed to have the form proposed by \cite{Flory42} and thus the total free energy associated with the impregnating solvent is
\begin{equation}\label{psim}
\psi_s\left(\phi^{\rm R}\right)=\phi^{\rm R}\left\{ \psi_0+ \frac{kT}{\nu}\left[\ln \left(1-\frac{1}{1+\phi^{\rm R}}\right)+\frac{\chi}{1+\phi^{\rm R}}\right]\right\},
\end{equation}
where $\chi$ represents the Flory-Huggins interaction parameter and $\nu$ the volume of a solvent unit.
Substituting the expression of the Helmholtz free energy  \eqref{psi} (with \eqref{psie} and \eqref{psim}) in \eqref{constitutiveCauchyPiola} and \eqref{constitutivemu}, and using the relation \eqref{phireference}, we  can  now write the   Cauchy and Piola stress tensors as
\begin{equation}\label{CauchySpecific}
{\bf T}=\frac{G}{J}\left[{\bf F}{\bf F}^{\rm T}-\left(1+J\frac{ p}{G}\right) {\bf I}\right],\qquad{\bf S}={G}\left[{\bf F}-\left({1}+J\frac{ p}{G}\right)\ {\bf F}^{\rm -T}\right],
\end{equation}
respectively, and the chemical potential as

\begin{equation}\label{muspecific}
\mu = \mu_0 + \frac{kT}{\nu}\left[\ln \left(1-\frac{1}{J}\right)+\frac{1}{J}+\frac{\chi}{J^2}\right]+p.
\end{equation}

The solvent diffusion within the solid matrix is assumed to follow the classical model \citep{Feynman}  that has the form
\begin{equation}\label{jspecific}
{\bf j}=-\frac{D\nu}{kT}\phi\ {\rm grad}\ \mu,
\end{equation}
where $D$ is the diffusion coefficient. The diffusion is hence proportional to the gradient of chemical potential  and to the solvent volume fraction. The referential and spatial gradients are related by ${\rm Grad}\ \mu={\bf F}^{\rm T}\ {\rm grad}\ \mu$. Thus the referential description of the kinetic law above corresponds to a mobility tensor, defined in \eqref{kineticlaw}, of the form
\begin{equation}\label{jRspecific}
 \quad {\bf M}\left({\bf F}, \phi^{\rm R} \right)={\frac{D\nu}{kT}}\phi^{\rm R}{\bf B}^{-1} \quad \text{where} \quad {\bf B}={\bf F}^{\rm T}{\bf F}.
\end{equation}
Note that ${\bf M}$ is symmetric and positive definite provided that  $D>0$.

A full description of the material response in the bulk is thus defined by a set of  material parameters $(\nu,N, \chi, \psi_0,D) $.  In solving a specific growth problem in the next section  we will use common values of the model parameters found in the literature: ${\nu} =10^{-28}{\rm m}^{3}$, $N=10^{24}{\rm m^{-3}}$, $\chi=0.2$ and $\psi_0=-4\times10^5 {\rm Jm^{-3}}$, and  assuming room temperature we have $kT=4\times 10^{-21} \rm J$. Investigation of the constitutive sensitivity of our results will center on model parameters that are associated with growth kinetics, and will thus include the influence of   diffusion rates that are dictated by the diffusion coefficient $D$.

\bigskip

\noindent\textbf{Constitutive relations on the boundary.} The latent energy of growth is defined as the jump of potential energy across the boundary \eqref{latentenergyDef}. On the inner side of the boundary (within the body), the free energy is  defined by equation \eqref{psi}. Considering an impermeable growth surface, association on this boundary  is promoted by a {chemical binding potential} that locally  alters the potential energy, such that $\psi^+=\psi_a$, and represents the energetic gain due to association. Therefore, the latent energy of association reads
\begin{equation}
 \Delta\psi= \psi_{a}-\psi. \label{Dpsia}
\end{equation}
  At the dissociation surface material on the inner side of the boundary (within the body)  has  free energy $\psi^-=\psi$ and a volume ratio of $J$; the same  volume of material on the outer side of the boundary (in the solvent region) has a free energy of $\psi^+=J\psi_0$. The latent energy at the dissociation surface is thus
\begin{equation}
\Delta\psi=J\psi_0-\psi. \label{Dpsid}
\end{equation}
Notice that in this case, we assume that the dissociation happens naturally and is not prompted by a local energy.

The remaining constitutive relation that needs to be defined is  the kinetic law of growth \eqref{fVRcondition}, or in other words, the growth function $ \mathcal{G}(f)$. Although the present study is concerned with modeling surface growth phenomena from a macroscopic point of view by applying  tools of continuum mechanics, the surface growth reaction occurs locally and is driven by a \textit{Brownian Ratchet} mechanism of polymerization/depolymerization; a process
governed by random walks and thermal fluctuations \citep{Mogilner96, Mogilner2003-Force, Mogilner2003-Motors,Theriot}. Nonetheless, by observing biological growth, it is apparent that    the stochastic nature of these processes at the microscopic level does not  transcend into the macroscopic response.      Here we apply the relation for the driving force \eqref{drivingforce} to    tie between the locally occurring  chemo-mechanical process,  and the deterministic response at the macroscopic level.

We consider a simplified framework in which chemical reactions take place only at the free ends of the polymer network that are exposed at the boundaries of the body. The intrinsic polarity of the solid matrix is such that the reaction rates are different at either end \citep{Theriot}. When considering polymerization, the rate of association is typically  referred to as $r_{\rm on}$ while the rate of dissociation as $r_{\rm off}$. The absolute rate of growth is thus regularly written in the form $r=r_{\rm on}-r_{\rm off}$.
Hence, for association we have $r>0$ while for dissociation $r<0$. We can relate these reaction rates with the energy barrier  that must be overcome to induce them, which is the work invested by the driving force in binding/unbinding a single unit, i.e. $\nu f$. Therefore, we can describe the thermally activated process by applying a growth function  of an Arrhenius form
\begin{equation}\label{VR}
\mathcal{G}(f)=\frac{b}{2}\left(e^{\frac{\nu f}{kT}}-e^{-\frac{\nu f}{kT}}\right)=b \sinh \left( \frac{\nu f}{kT}\right),
\end{equation}
that obeys the thermodynamic inequality \eqref{fVpositive} for a  reaction constant $b>0$, which according to \eqref{fVRcondition}, scales with the growth rate. Here, each exponential corresponds to a different reaction. The first exponential represents $r_{\rm on}$ and the second   exponential represents $r_{\rm off}$.  In the limit of small departures from thermodynamic equilibrium, \eqref{VR} reduces to the linear form

\begin{equation}\label{VRlinear}
\mathcal{G}(f)=b  \left( \frac{\nu f}{kT}\right).
\end{equation}

In the present study we choose the kinetic law of growth to be  identical for both the association and dissociation surfaces and introduce  the polarity  through the latent energy of growth \eqref{Dpsia} and \eqref{Dpsid}. This formulation allows the growth surfaces to smoothly transition from association to dissociation and vice versa if it becomes energetically favorable.

\section{A boundary-value problem: {Growth on a flat surface}}\label{1Dproblem}
\noindent The theoretical framework presented in the previous sections can be applied to study coupled growth processes occurring in an arbitrary geometry. In this section, we demonstrate its application for the specific problem of growth on a flat substrate. It will be shown that this simple growth setting generates nontrivial insights into the kinetics of surface growth with coupled diffusion. Moreover, it provides a good approximation for growth of thin layers on surfaces of arbitrary geometry.

\begin{figure}[ht]
\begin{center}\includegraphics[width=0.7\textwidth]{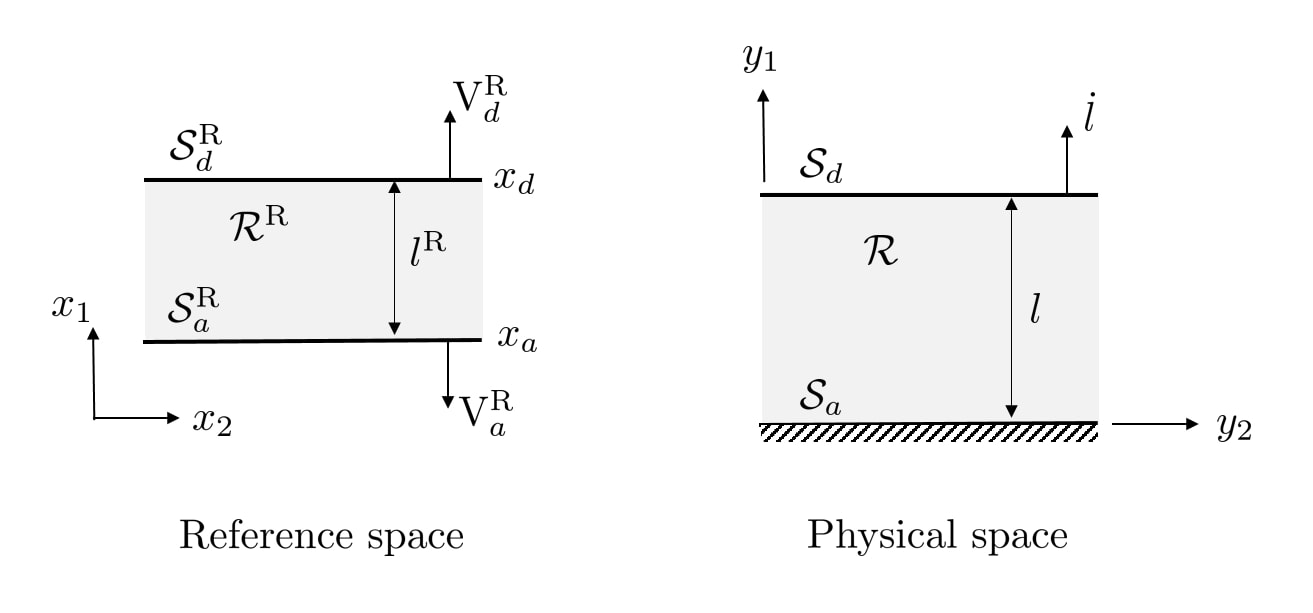} \end{center}
\caption{Continuum representation of a body grown on an infinite flat surface in the reference space and in the physical space.}
\label{1D}
\end{figure}

As for the derivation of the general model, in solution of the specific boundary value problem we make use of both the physical and referential spaces. The physical space is described by the set of axis $(y_1, y_2,y_3)$ where the $y_1$-axis is perpendicular to the rigid substrate. At a given time $t$, the  body manifold occupies the region
\begin{equation}
0\le y_1 \le l(t), \quad -\infty < y_2,y_3 < +\infty,
\end{equation}
where $l(t)$ is the thickness of the body (Fig. \ref{1D}). Association occurs on the substrate  $y_1=0$ and dissociation occurs on the free surface  $y_1=l(t)$.

The reference space is described by the set of axis $(x_1, x_2,x_3)$ where the $x_1$-axis is perpendicular to the association surface. At a given time $t$, the material manifold occupies the region
\begin{equation}
x_a(t)\le x_1 \le x_d(t), \quad -\infty < x_2,x_3 < +\infty,
\end{equation}
where $x_1=x_a(t)$ and $x_1=x_d(t)$ represent the association and dissociation surfaces respectively.

We consider a situation in which growth is promoted on the substrate by a homogeneous  density of binding sites that imposes a constant equibiaxial in-plane stretch  $\lambda _0$ of the grown material.
The mapping  from the reference to the current configuration \eqref{position} thus reduces to the form
\begin{equation}\label{1Dposition}
 y_1=\hat {y}_1(x_1,t), \quad y_2=\lambda_0 x_2,\quad y_3=\lambda_0 x_3,
\end{equation}and the swelling ratio can be expressed as

\begin{equation}\label{1DJ}
J=\lambda \lambda_0^2,
\end{equation}where  $\lambda$ denotes the out-of plane stretch
\begin{equation}\label{1DLambda}
\lambda(x_{1},t)=\frac{\partial \hat y_1}{\partial  x_1}.
\end{equation}For future use we denote the swelling ratios at the association and dissociation boundaries by    $J_a=J\left(y_1=0,t\right)$  and $J_d=J\left(y_1=l,t\right) $, respectively.

 Since all of the motion occurs along the growth direction, the material velocity and flux can be written as ${{\bf v}={\rm v}(x_1,t){\bf e}_{y_1}}$ and ${\bf j}={\rm j}(x_1,t){\bf e}_{y_1}$ where

\begin{equation}
{\rm v}(x_{1},t)=\frac{\partial \hat y_1}{\partial t}.
\end{equation}In this setting, the velocity of the association and dissociation  surfaces in the current configuration are
\begin{equation}\label{1DV}
{\bf V}={\bf 0} \quad \text{and}\quad {\bf V}=\dot l  {\bf e}_{y_1},
\end{equation}
respectively. Similarly  in the reference configuration the boundary velocities are
\begin{equation}\label{1DVR}
{\bf V}^{\rm R}=-{\rm V}^{\rm R}_a{\bf e}_{x_1}\quad \text{and}\quad  {\bf V}^{\rm R} ={\rm V}^{\rm R}_d{\bf e}_{x_1},
\end{equation}where ${\rm V}^{\rm R}_a$ and ${\rm V}^{\rm R}_d$ denote the referential association and dissociation velocities respectively.

Now, let $l^{\rm R}(t)$ denote the thickness of the growing layer in the reference frame, that we will refer to as the \textit{dry thickness}, its time derivative can be expressed in terms of association and dissociation rates as

\begin{equation}\label{lRVaVd}
\dot l^{\rm R}={\rm V}^{\rm R}_a+{\rm V}^{\rm R}_d.
\end{equation}

Turning to considerations of mass conservation, we combine conservation equation \eqref{vjnur} with the jump condition \eqref{jin} on the impermeable association surface, to relate the  material velocity and the solvent flux by
\begin{equation}
{\rm v}+{\rm j}=0, \label{velocityflux}
\end{equation}
and equation \eqref{dJdt} specialized to this uniaxial problem yields
\begin{equation}\label{diffusioneq}
\frac {\partial J}{\partial t}=J^2 \frac{\partial}{\partial y_1}\left(\frac {\rm v}{J}\right).
\end{equation}

Employing the constitutive relation \eqref{CauchySpecific} we can obtain the principal Cauchy stress components
\begin{equation}\label{sigma12}
{\rm T}_{11}=NkT\left( \frac{J}{\lambda_0^4}-\frac{1}{J} \right)-p, \quad \quad
{\rm T}_{22}=NkT\left( \frac{\lambda_0^2}{J}-\frac{1}{J} \right)-p.
\end{equation}

In the absence of body forces, mechanical equilibrium \eqref{Cauchy} implies that the out-of-plane stress vanishes through the thickness of the body $({\rm T}_{11}=0)$, hence the hydrostatic pressure as a function of the swelling ratio is
\begin{equation}\label{pressure}
p\left(J\right)=NkT\left(\frac{J}{\lambda_0^4}-\frac{1}{J}\right).
\end{equation}
The in-plane stress component \eqref{sigma12}$^2$ is non-zero and can now be written as

\begin{equation}\label{sigma2nopressure}
{\rm T}_{22}=NkT\left( \frac{\lambda_0^2}{J}-\frac{J}{\lambda_0^4} \right),
\end{equation}
and similarly the chemical potential \eqref{muspecific} reads

\begin{equation}\label{muJ}
\mu(J) = \mu_0 + \frac{kT}{\nu}\left[\ln \left(1-\frac{1}{J}\right)+\frac{1}{J}+\frac{\chi}{J^2}\right]+NkT\left(\frac{J}{\lambda_0^4}-\frac{1}{J}\right).
\end{equation}
Using to \eqref{phireference} to relate $\phi^{\rm R}$ to $J$, and \eqref{1Dposition} and \eqref{1DJ} to relate the components of $\bf F$ to $J$,  the free energy given in \eqref{psi}, \eqref{psie} and \eqref{psim} can also be expressed as a function of $J$ as follow
\begin{equation}
\psi(J)=\frac{NkT}{2}\left(\frac{J^2}{\lambda_0^4}+2\lambda_0^2-3-2 \ln  J\right)+(J-1)\left\{ \psi_0+ \frac{kT}{\nu}\left[\ln \left(1-\frac{1}{J}\right)+\frac{\chi}{J}\right]\right\}.
\end{equation}
As a result, the driving force at the boundary \eqref{drivingforce} can be written as a function of $J$  in the form
\begin{equation}\label{fJ}
f\left(J\right)= \Delta \psi\left(J\right) +\mu\left(J\right) {J}.
\end{equation}  Corresponding relations for the driving force at the association and dissociation boundaries     $f(J_a)$ and  $f(J_d)$, respectively, are found in \eqref{fasteadyAppendix} and \eqref{fdsteadyAppendix} of  Appendix B. Consequently, by the kinetic law \eqref{VR}, the association and dissociation rates  are also  functions of $J$, namely ${\rm V}^{\rm R}_a={\rm V}^{\rm R}_a\left(J_a\right)$ and ${\rm V}^{\rm R}_d={\rm V}^{\rm R}_d\left(J_d\right)$.

Now, using the  diffusion  law \eqref{jspecific} and relation \eqref{velocityflux}, we can express the ratio ${\rm v}/J$ as
\begin{equation}\label{vJrelation}
\frac{\rm v}{J}=\Lambda\left(J\right)\frac{\partial J}{\partial y_1}\quad \text{where}\quad  \Lambda (J)=\frac{D\nu}{kT}\left(\frac{J-1}{J^2}\right)\frac{d\mu }{dJ}.
\end{equation}
By substituting  the specific expression of the chemical potential \eqref{muJ} we can write

\begin{equation}\label{gfunction}
\Lambda\left(J\right)=D\left[\frac{N\nu}{\lambda_0^4}\frac{1}{J}-\frac{N\nu}
{\lambda_0^4}\frac{1}{J^2}+{N\nu}\frac{1}{J^3}+\left(1-{N\nu}-2\chi\right)\frac{1}{J^4}+2\chi \frac{1}{J^5}\right].
\end{equation}

The combination of the two equations \eqref{diffusioneq} and \eqref{vJrelation} results in a  single partial differential equation for the field variable $J(y_1,t)$, which is sufficient to determine the response in the bulk.
To complete the representation of the boundary value problem, it remains to determine the values of $J$ on the association and dissociation surfaces. The coupling between surface growth and swelling is generated  by the dependence of the association and dissociation reactions on the local values of the field  $J$. To formulate this dependence, let us first examine the dissociation boundary. The continuity of chemical potential \eqref{muJ} at $y_1= l(t)$ implies
\begin{equation}\label{potentialBC}
\mu\left(J_d \right)=\mu_0,
\end{equation}
thus resulting in an implicit relation for $J_d$ (provided in full form  \eqref{JdsteadyAppendix} in Appendix B). Holding the reference chemical potential of the surrounding solvent $\mu_0$ constant,  $J_d$ is also constant in time, and is dictated by material parameters ($\chi$, $N$, $\nu$), and the in-plane bi-axial stretch $\lambda_0$.

By the kinematic relation \eqref{velocity}, the particle velocity at the dissociation boundary is

\begin{equation}\label{vdBC}
{\rm v}_d=\dot l-\frac{J_d}{\lambda_0^2}{\rm V}^{\rm R}_d  (J_d),
\end{equation}
notice that the dissociation rate ${\rm V}^{\rm R}_d={\rm V}^{\rm R}_d \left(J_d \right)$  is constant in time.

Now let us examine the association boundary. The material velocity  ${\rm  v}_a$ at the interface can be written with the aid of the kinematic relation \eqref{velocity} as a function of  $J_a$
\begin{equation}\label{vaBC}
{\rm v}_a=\frac{J_a}{\lambda_0^2}{\rm V}^{\rm R}_a \left(J_a\right).
\end{equation}
The combination of \eqref{vJrelation} and \eqref{vaBC} implies a boundary condition on the association surface that couples the kinetics of association to the swelling ratio
by\begin{equation}\label{vaBC2}
\Lambda (J)\frac{\partial J}{\partial y_1} -\frac{ { {\rm V}^{\rm R}}_a \left(J_a\right)} {\lambda_0^2}=0 \qquad \text{at} \qquad{y_1=0}.
\end{equation}

Finally, the governing equations can be summarized in the physical space by the boundary value problem
\begin{equation}\label{systemphysical}
\begin{cases}
{\frac {\partial J}{\partial t}=J^2 \frac{\partial}{\partial y_1}\left(\Lambda\left(J\right)\frac{\partial J}{\partial y_1}\right)},  &0\le y_1 \le l(t)
\\
\left[\![ \mu\left(J \right) \right ]\!] =0, &{y_1=l(t)}
\\
\Lambda\left(J\right)\frac{\partial J}{\partial y_1}-\frac{{\rm V}^{\rm R}_a\left(J\right)}{\lambda_0^2}  =0, &{y_1=0}
\\
\end{cases}
\end{equation}
where $l(t)$ is given by \eqref{vdBC}. Alternatively, we can write the problem in the reference space with the following equations\begin{equation}\label{systemreference}
\begin{cases}
\dot J=\frac{\partial}{\partial x_1}\left(\lambda_0^4 \Lambda\left(J\right)\frac{\partial J}{\partial x_1} \right), &x_a(t)\le x_1 \le x_d(t)
\\
\left[\![ \mu\left(J \right)\right ]\!] =0, &{x_1=x_d(t)}
\\
\lambda_0^4 \Lambda \left(J\right)\frac{\partial J }{\partial x_1}-J{\rm V}^{\rm R}_a(J)=0, &{x_1=x_a(t)}
\end{cases}
\end{equation}
where  $J$ refers to the swelling ratio expressed either in the reference space coordinates or the physical space coordinates indifferently. Hence, we have reduced the problem to be fully defined by a single field variable, the swelling ratio, and the time dependent thickness of the layer, which can be written in the current or reference configurations, as  $(l(t),J(y_1,t))$  or  $(l^{\rm R}(t),J(x_1,t))$, respectively. Once the field $J$ is known, all other fields  (deformation, stresses, chemical potential) can be identified.

While both of the representations are equivalent, a specific advantage of the reference frame is in allowing us to separate the two species, since the dry thickness  $l^{\rm R}(t)$ indicates the amount of solid   solely. Nonetheless, in the following sections, we will alternate between both configurations.

Although the considered growth process is highly nonlinear, to obtain a better understanding of the time and length scales associated with growth we observe that from the  differential equation of \eqref{systemreference} with \eqref{gfunction}, the diffusion rate  in  gels with finite swelling scales with   $(DN\nu)/L$, where  $L$  is the characteristic dry length of the system.  On the other hand, according to \eqref{VR} the  growth rate scales with $b$. However, from the second boundary condition in  \eqref{systemreference} along with the identity \eqref{Jphi}, it is evident that these rates are not independent and    the ratio between them is ultimately dictated by the local swelling gradients that are generated by the growth reaction\begin{equation}\label{lc}
\frac{\partial \phi^{\rm R}\ }{\partial \xi_1}\propto \frac{bL}{DN\nu}=\frac{L}{l_{c}},
\end{equation}where $\xi_1=x_1/L$ and the critical length scale is defined as   $l_c=DN\nu/b$. In other words,  if the layer thickness is much smaller than the critical length  $(L\ll l_c),$ the swelling gradients  are small thus suppressing diffusion, and if the layer thickness is much larger than the critical length   $(L\gg l_c),$  swelling gradients at the association boundary accelerate the diffusion to sustain the growth. Hence,  the rate coefficients (i.e. $D$ and $b$) dictate the critical length scale, and determine the local gradients of swelling needed to sustain the growth. However, since the material is growing, the characteristic length scale of the system can significantly change throughout the process and the time scales vary accordingly.

\bigskip
\noindent \textbf{{Treadmilling response.}}\label{Treadmilling}
If  addition and removal of mass are balanced, the system will arrive at a treadmilling regime, a steady state in which the layer thickness is constant although association and dissociation are continuously occurring. Once the system has arrived at this the treadmilling state, all time derivatives  in the current configuration vanish. Therefore, the boundary value problem \eqref{systemphysical} can be solved  analytically to determine the steady thickness $\tilde l$ and the steady swelling ratio field $\tilde J(y_1)$, where the superscript $\tilde {( \ )}$ denotes treadmilling values.

According to \eqref{lRVaVd}, once   addition and removal of solid mass are balanced, we can write
\begin{equation} \label{Jainf}
{\rm V}^{\rm R}_a(\tilde J_{a})=-{\rm V}^{\rm R}_d (\tilde J_d),
\end{equation}
where $\tilde J_{a}=\tilde J(y_1=0)$ denotes the steady swelling ratio  at the association boundary. Since ${\rm V}_d^{\rm R}$ is constant,  $\tilde J_{a}$ can be determined from  \eqref{Jainf} with  ${\rm V}_a^{\rm R}$ and  ${\rm V}_d^{\rm R}$   related to the driving force \eqref{fJ} by the kinetic relation  \eqref{VR}.

 Combining \eqref{Jainf} with \eqref{systemphysical}, in absence of time dependence,  a differential equation for the steady field $\tilde J$ can be obtained
\begin{equation}\label{gammaconstant}
\Lambda(\tilde J)\frac{d \tilde J}{d y_1}=\frac{{\rm V}^{\rm R}_a(\tilde J_{a})}{\lambda_0^2},
\end{equation}
which by integration yields an implicit relation between the  swelling ratio $\tilde J$ and the spatial coordinate $y_1$
\begin{equation}\label{y-stationary}
y_1 (\tilde J )=\frac{\lambda_0^2}{{\rm V}^{\rm R}_a(\tilde J_{a})}\int_ {\tilde J_a}^{\tilde J}\Lambda(J)d J.
\end{equation}
The stationary thickness in the physical space can thus be derived as
\begin{equation}\label{l-stationary}
\tilde l=y_{1}(J_d).
\end{equation} Although in the reference frame the boundaries of the body at treadmilling move at  a  constant velocity given by \eqref{Jainf}, the distance between any two spatial points mapped back into the reference frame remains  constant, due to self-similarity. Therefore, employing the inverse transformation $x_1=\hat x_1(y_1,t)$, we  define the distance from the association surface in the reference frame as $\Delta x_1=  \hat x_1(y_1,t)-\hat x_{1}(0,t)$ which   by  integration of  the  transformation  \eqref{1DLambda} reads
\begin{equation}\label{x-stationary-reference frame}
\Delta x_1=\int_{0}^{ y_1}\frac{\lambda_0^2}{\tilde J(y_1)}{dy_1.}
\end{equation}
Now we substitute equation   \eqref{gammaconstant}
to obtain an implicit relation between the steady field $\tilde J$ and  the distance $\Delta x_1$
\begin{equation}\label{Deltax}
\Delta x_1(\tilde J)=\frac{\lambda_0^4}{{\rm V}^{\rm R}_a(\tilde J_{a})}\int_ {\tilde J_{a}}^{\tilde J}\frac{\Lambda(J)}{J}dJ.
\end{equation}
The dry thickness $\tilde l^{\rm R}$ is then given  by
\begin{equation}\label{l-stationary-reference}
\tilde l^{\rm R}=\Delta x_1(J_d).
\end{equation}
{Analytical relations for $y_1 (\tilde J )$ and $\Delta x_1 (\tilde J )$ are given in relations \eqref{Jy1steady} and \eqref{Jx1steady} of Appendix B.}

\bigskip

\noindent {\bf Integration results and parameter sensitivity.} Given the above formulation, the steady state, if reached, is unique and characterized by $(\tilde l, \tilde J(y_1))$ in the physical frame, or alternatively by $(\tilde l^{\rm R}, \tilde J(\Delta x_1))$ in the reference frame. To study the  effect  of model parameters that are associated with the growth kinetics on the treadmilling response, we hold the material parameters defined in the previous section   $(\nu$, $N$,  $\chi$, $\psi_0$)  constant, and examine the variation of $(\tilde l^{\rm R}, \tilde J(\Delta x_1))$ with the kinetic parameters, namely the diffusion coefficient - $D$, the reaction constant - $b$ and the  potential energy gain at the association surface -  $\psi_a$.

First, from equations \eqref{Deltax} and \eqref{l-stationary-reference}, along with the kinetic law \eqref{VR}, it is observed that ${\rm V}^{\rm R}_d$ is proportional to $b$  while  $\Lambda(J)$ is proportional to $DN\nu$ (by its definition in equation \eqref{gfunction}), hence the lengths   $\Delta x_1$ and $\tilde l^{\rm R}$ scale with the critical length   $l_{c}$ defined in \eqref{lc}. On the other hand, the swelling ratio    $\tilde J_{a}$ is not affected by the rate coefficients ($D$ and $b$), as observed from  equation \eqref{Jainf}, nor is the distribution of the  treadmilling swelling ratio      $\tilde J$. In the following sections  we take $DN\nu=10^{-8}{\rm m^2s^{-1}}$ and $b=10^{-7}\rm ms^{-1}$; the critical length is thus $l_c=10^{-1}\rm m$.
\begin{figure}[ht]
\begin{center}\includegraphics[width=1.0\textwidth]{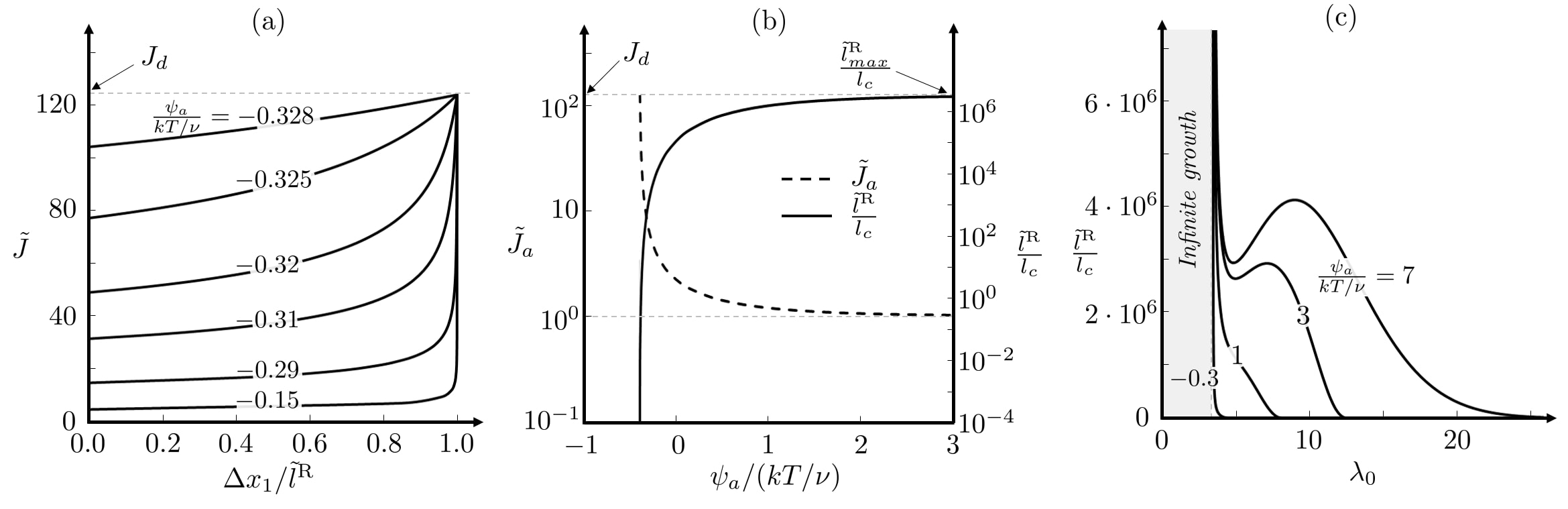} \end{center}
\caption{Sensitivity of treadmilling response to growth parameters: (a) Distribution of swelling through the layer  $(\Delta x_1/\tilde l^{\rm R}) $  for various values of the potential energy gain at the association surface $\psi_a$ with $\lambda_0=4.96$. (b) Swelling ratio at the association surface $\tilde J_a$ and the dimensionless   thickness $\tilde l^{\rm  R} / l_{c}$ as a function of the dimensionless  potential  energy gain $\psi_a/(kT/\nu)$, with $\lambda_0=4.96$. Logarithmic scales  are used to present the  swelling ratio and the dimensionless  thickness (c) Dry thickness as a function of the in-plane stretch at the association boundary $\lambda_0$ for various values of $\psi_a$. }
\label{PsiATread}
\end{figure}

The  potential energy gain at the association boundary $\psi_a$ has a more complex effect on the treadmilling response. In Fig. \ref{PsiATread}(a) we show curves for the treadmilling swelling ratio as a function of the normalized distance    from the association surface in the reference frame $\Delta x_1/\tilde l^{\rm R}$, for various values of  $\psi_a/(kT/\nu)$.  It is observed that as $\psi_a$ increases the swelling ratio at the association boundary decreases and swelling gradients  localize near the dissociation boundary. From the limited range of  $\psi_a$ values in Fig. \ref{PsiATread}(a), it is apparent  that the energy $\psi_a$ has to fall within a certain range for growth to be both physically possible and energetically favorable. First the steady dry thickness needs to have a positive value $\tilde l^{\rm R} \ge 0$, which, by combination of \eqref{Deltax} and \eqref{l-stationary-reference}, requires $J$ to be an increasing function thus leading to the inequality  $ \tilde J_a \le J_d$, that defines a lower bound limit such that  $\psi_{a}\ge -0.329(kT/\nu) $, as derived from equation \eqref{Jainf}. Although there is no physical upper bound limit on the value of  $\psi_a$, it is found that a  $\psi_a$ increases beyond the range presented in Fig.  \ref{PsiATread}(a) changes become insignificant while the swelling ratio at the association boundary tends to the incompressibility limit $\tilde J_a\to1$. Hence,  beyond a certain limit the treamilling response becomes insensitive to $\psi_a$. This point is further explained by the curves presented in Fig. \ref{PsiATread}(b), where we show the dependence of the association swelling ratio  $\tilde J_a$  and the dimensionless treadmilling length  $\tilde l^{\rm  R}/l_{c}$ on $\psi_a/(kT/\nu)$. First we notice that when $\psi_a$ approaches its lower bound, the thickness of the layer vanishes such that $\tilde l^{\rm R} \to 0$  and $\tilde J_a \to J_d$. On the other hand, as $\psi_a$ increases, $\tilde J_{a}\to1$, and the thickness of the layer approaches an asymptotic limit  $\tilde l^{\rm R}\to\tilde l^{\rm R}_{max}$ (which can be determined analytically via \eqref{Deltax} and \eqref{l-stationary-reference} with $\tilde J_a=1)$.  Physically, a higher association energy enhances the association reaction by kinetic law \eqref{VR}, thus leading to a larger steady dry thickness, however this becomes inefficient as the  solid matrix  becomes more dense thus resisting the flux of solvent that is needed to generate growth.  {At the limit  $     (\psi_a\to \infty)  $ solvent cannot reach the association surface thus choking the growth.

Next  we explore the sensitivity of  $\tilde l^{\rm R}$ to  the in-plane stretch $\lambda_0$ that is dictated by the properties of the growth surface. It is worth emphasizing that for a given set of material parameters,  $\lambda_0$ is the {only} tuneable parameter that can be adjusted to determine the growth. In natural systems,  $\lambda_0$ is controlled by the distribution of binding sites on the growth surface  (recall that   $1/\lambda_0$  is the  fraction of solid volume generated on the growth surface). From Fig. \ref{PsiATread}(c)   it is immediately noticed that   $\lambda_0$  has a nonintuitive effect on the growth. If the  growth reaction is   sparse and  $\lambda_0\gg10 $   then dissociation becomes
 energetically favorable and the thickness of the layer vanishes. At the other end, if the {growth reaction is dense }   $(\lambda_0< 3.5)$,  this leads to smaller values of $J_d$, which by \eqref{fJ}  result in a change of sign of the driving force  so that  it becomes energetically favorable to associate at the free boundary. Hence, association occurs on both boundaries and the layer will continue to grow indefinitely. In the intermediate zone $(\lambda_0> 3.5)$ the rate of dissociation increases monotonically with increasing   $\lambda_0$, thus contributing to the  reduction in maximum thickness, however we observe a non-monotonic dependence on{  $  \lambda_0$ as  $\psi_a$ increases: first a local minimum appears then a local maximum is observed}.} {This behaviour is due to the competition between two effects: increasing values of $\psi_a$ favor growth, while increasing values of $\lambda_0$ hinder it.} To further understand this complex response it is essential to consider the evolution of the growth before it has arrived at the treadmilling state.

\bigskip
\noindent\textbf{{Numerical results and the emergence of a universal path.}}
 Although it has been established in the previous section that, under certain conditions, a treadmilling response can exist, it has yet to be determined whether and how the  boundary value problem, defined by \eqref{systemphysical} and \eqref{systemreference}, evolves towards this treadmilling state. Is this evolution sensitive to initial conditions? And more importantly, how does a layer evolve from an initial state with no solid at all? In other terms, can this model capture the entire evolution of the body from inception up to treadmilling?

 An added complexity of the present problem arises due to the existence of  moving boundaries in both the reference and the physical space. Integration of the boundary value problem defined in either \eqref{systemphysical} or \eqref{systemreference}, is thus performed via a specialized numerical method that incorporates two integration time scales, to track both the moving boundaries and the diffusion mechanism.
 In the following,  we choose to integrate the boundary value problem in the reference frame to  examine the solid matrix and the solvent separately.
We solve for  $(l^{\rm R}(t),J(x_1,t))$ in order to  describe the time evolution of the system for various initial conditions $(l^{\rm R}(0),J(x_1,0))$.

\begin{figure}[ht]
\begin{center}\includegraphics[width=0.95\textwidth]{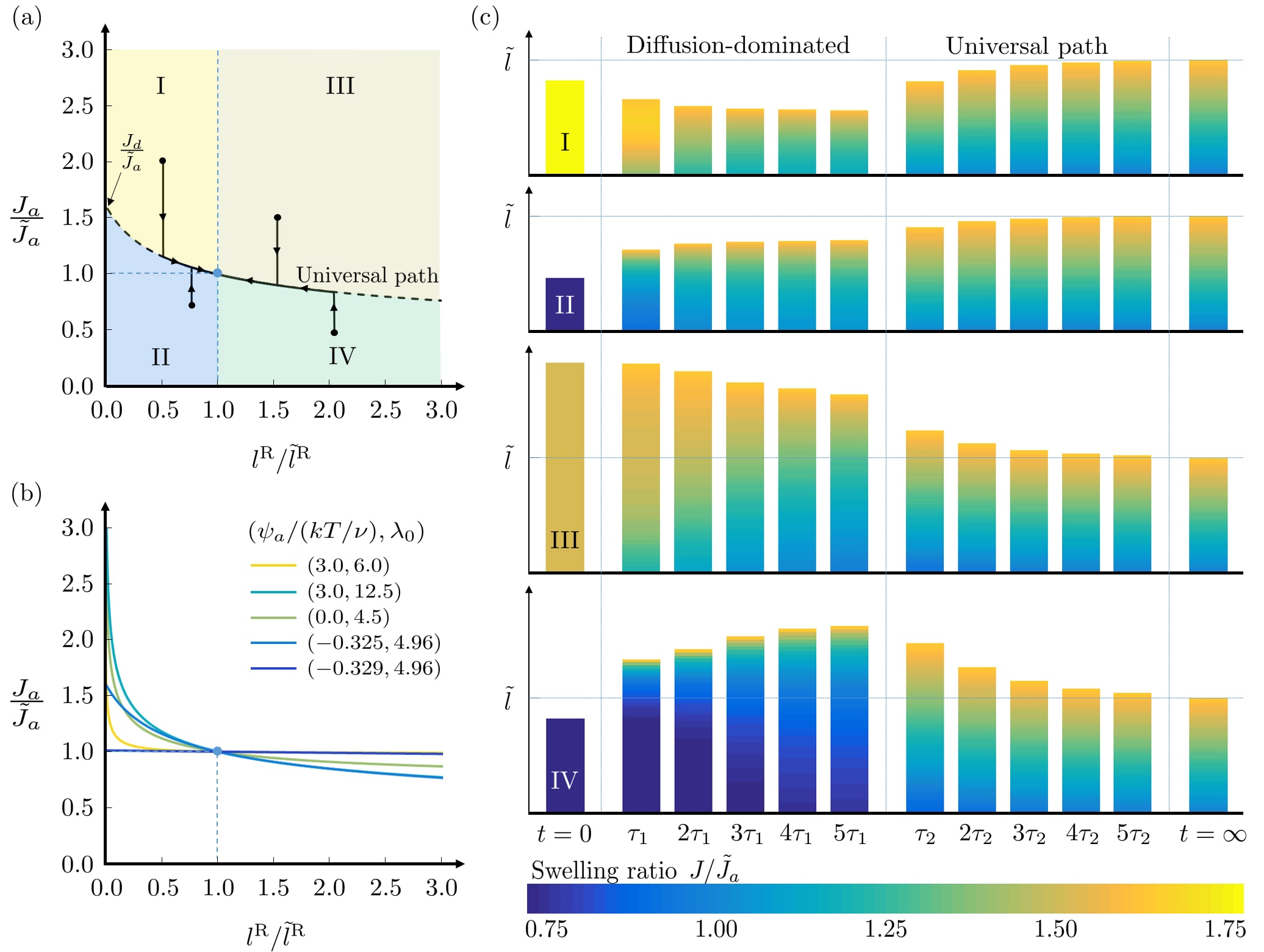} \end{center}
\caption{(a) Evolution of a growing system represented in the phase space  of normalized association swelling ratio  versus normalized dry thickness, with $\psi_a/(kT/\nu)=-0.325$ and $\lambda_0=4.96$.  Four regions (I - IV) represent different regimes of growth response. A growth path initiating from each region is shown by the black lines with the arrows representing the direction of the evolution. (b) Effect of  $\psi_a$ and  $\lambda_0$ on  the universal path represented in the normalized phase space. The universal paths presented in (a) and (b)  derived from numerical simulations and analytical formula \eqref{UniversalRelation},  and  are indistinguishable.  (c) Time evolution of total thickness in the physical space and swelling ratio throughout the thickness for the specific paths starting in regions I, II, III and IV. }
\label{Phase}
\end{figure}

In Fig. \ref{Phase}(a), we plot the evolution of the growing body from an initially homogenous layer of given dry thickness    $l^{\rm R}$ and spatially constant swelling ratio $J$ (indicated by the black dots) towards the treadmilling state (indicated by the blue dot). Given the importance of $J_a$ and its crucial role in driving the coupled growth,  results are represented in  a phase space defined by the   dimensionless dry thickness of the body $l^{\rm R}(t)/\tilde l^{\rm R}$ and the renormalized swelling ratio ${J_a(t)}/{\tilde J_{a}}$. The arrows indicate the direction of the evolution toward treadmilling  $(l^{\rm R}(t)/\tilde l^{\rm R},{J_a(t)}/{\tilde J_{a}})=(1,1)$.
On this note, it is worth emphasising  that  the treadmilling limit recovered by  space-time integrations agrees with the  analytically derived limit in \eqref{l-stationary-reference}.

Two distinct {regimes} clearly emerge from the evolution paths in Fig. \ref{Phase}(a): first, a \textit{diffusion-dominated regime} during which changes in dry thickness are relatively small, then following a sharp change in trend, a second regime in which surface growth and solvent diffusion are fully coupled. In this {regime} the system follows a path that is independent of initial conditions and is thus referred to as the \textit{universal path}.

Although only four paths are represented in Fig. \ref{Phase}(a), a comprehensive analysis of the sensitivity to initial conditions (including nonhomogeneous initial states)  has been conducted and  a similar sharp transition from a diffusion-dominated regime to the  universal path   has been observed in all cases and across a range of constitutive parameters (Fig. \ref{Phase}(b) shows the universal path for various combinations of $\psi_a$ and $\lambda_0$). Nonetheless we identify four different regions in the phase diagram (as indicated by the roman numerals) that are characterized by distinct evolution scenarios. To explain these differences we present in Fig. \ref{Phase}(c) the time dependent evolution of the layers  corresponding to the paths represented in Fig. \ref{Phase}(a), each of which initiates in a different region.  The physical state of the system   at various times is shown  via a series of vertical bars that  indicate the physical thickness $l(t)$ and with color variation representing  the swelling ratio throughout the thickness $J(y_1,t)$.  The response in each of the two regime occurs at a different time-scale, hence results are represented first with time scale  $\tau_1$ of the  diffusion-dominated regime, then with time scale  $\tau_2$ of the evolution along the universal path. As previously mentioned, due to the highly nonlinear nature of the considered growth problem, these time scales are dependent on both the model parameters and initial conditions, and are not amenable to simple scaling arguments. To clearly present the results in the different regimes we thus normalize the time with respect to the representative  values $\tau_1=8.2\times 10^3\rm s$  and $\tau_2=1.6\times 10^7\rm s$. Examining  Fig. \ref{Phase}(c), we observe  non-monotonic evolution of the physical thickness of the body.  Layers that initiate in regions   I and III are over swollen, and thus release solvent  during the diffusion-dominated regime, while those that initiate in regions  II and IV are under swollen, and thus intake solvent. Once the system has reached the universal path, growth is driven by the coupling between surface growth and diffusion while swelling profiles maintain a nearly self-similar distribution that moves due to the local mechanisms on the boundaries. In this regime changes in dry thickness become significant and occur over a larger time scale. Systems initiating in regions I and II are under grown, and thus undergo an addition of solid mass during evolution along the universal path, while those initiating in region III and IV are over grown, and thus undergo removal of solid mass.

With the identification of the two distinct regimes, based on the numerical analysis, we now return to the analytical formulations to further characterize them by considering the dominant mechanisms that drive each of them.

\newpage
\noindent{\bf Diffusion-dominated regime.}
In this regime, the body undergoes rapid variations in  swelling to adjust to its environment, while changes in dry thickness (i.e. surface growth) is insignificant, as shown in Fig. \ref{Phase}(a). The bulk diffusion represented by the differential equation in  \eqref{systemreference}  is thus dominant and defines the time scale of evolution  while the time-dependent boundary conditions do not play a significant role, thus explaining the surprisingly  straight vertical lines that characterize the diffusion dominated regime  in the phase space. The time scale $\tau_1$ associated with this phase is thus proportional to  $L^{2}/(DN\nu)$, where $L$ is a characteristic dry length scale of the system.

\begin{figure}[ht]
\begin{center}\includegraphics[width=0.4\textwidth]{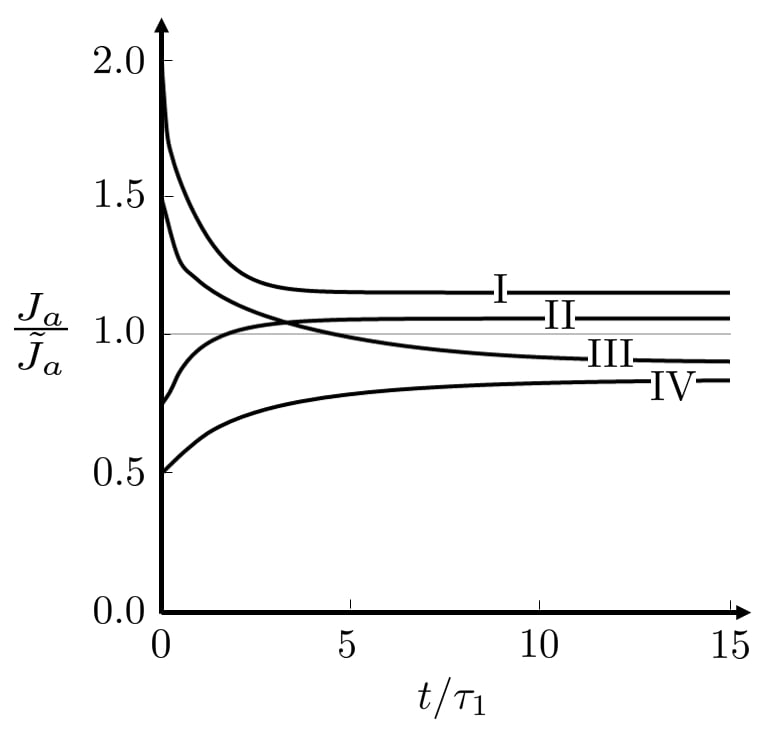} \end{center}
\caption{Time evolution of the normalized swelling ratio at the association boundary, during the diffusion-dominated regime for the paths initiating in regions I-IV and presented in Fig. \ref{Phase}(a). Time is normalized by $\tau_1=8.2 \times  10^3 {\rm s}$.}
\label{Tau1Swelling}
\end{figure}

Fig. \ref{Tau1Swelling} shows the evolution of the swelling ratio during  the diffusion-dominated regime for the four paths presented in Fig. \ref{Phase}(a).  In all cases, the swelling ratio at the association boundary tends to a nearly   asymptotic value as the system transitions to the   universal path.

\smallskip
\noindent{\bf The universal path.} After the transient variations of   swelling in the diffusion-dominated regime, the system arrives at a quasi-equilibrated state (Fig. \ref{Tau1Swelling}) in which surface growth and diffusion act harmoniously. This coupled growth is  governed by longer time scales hence  the time derivative on the left hand side of \eqref{systemphysical} becomes negligible, while the coupling between diffusion and surface growth  is manifested by the  boundary condition of \eqref{systemphysical}. The system hence simplifies to
\begin{equation}\label{universaldrive}
\Lambda\left(J\right)\frac{d J}{d y_1}=  \frac{{\rm  V}_a^{\rm   R}(J_a)}{\lambda_0^2}.
\end{equation}
Notice that this relation is similar to the expression for the treadmilling regime \eqref{gammaconstant} with a different that  the association rate,  and thus varies with time as $J_a$ evolves. {In a similar manner to the approach adopted for the treadmilling response, integration of \eqref{universaldrive} yields an implicit relation between the  swelling ratio $J$ and the spatial coordinate $y_1$
\begin{equation}\label{y-universal}
y_1 (J )=\frac{\lambda_0^2}{{\rm V}^{\rm R}_a(J_{a})}\int_ { J_a}^{J}\Lambda(J)d J,
\end{equation}
and the thickness in the physical space can thus be derived as $\tilde l=y_{1}(J_d)$.
By  integration of  the  transformation  \eqref{1DLambda}, and  substituting equation \eqref{universaldrive}, the distance from the association surface in the reference frame reads}

\begin{equation}\label{Deltax-universal}
\Delta x_1(J)=\frac{\lambda_0^4}{{\rm V}^{\rm R}_a(J_{a})}\int_ {J_{a}}^{ J}\frac{\Lambda(J)}{J}dJ.
\end{equation}
{Notice that functions $y_1(J)$ and $\Delta x_1(J)$ derived in the above equations are the same as the ones obtained in \eqref{y-stationary} and \eqref{Deltax}, with the difference that the swelling ratio has the superscript $\tilde{(\ )}$ for the treadmilling regime. Analytical expressions of the implicit spatial distribution of swelling are given in equations \eqref{Jy1steady} and \eqref{Jx1steady} of Appendix B.}

Finally, during the evolution along the universal path, the dry thickness $l^{\rm R}$ is  given by

\begin{equation}\label{UniversalRelation}
l^{\rm R}(J_{a})=\frac{\lambda_0^4}{{\rm V}^{\rm R}_a\left(J_{a}\right)}\int_ {J_{a}}^{J_d}\frac{\Lambda(J)}{J}dJ.
\end{equation}
This relation between $l^{\rm R}$ and $J_a$ provides an analytical form  for the universal   path shown in  Figs. \ref{Phase}(a,b). By comparing  with numerical results for various initial conditions we find that this analytical form agrees   with the numerically derived universal path.

Since the variation of dry thickness  $(l^{\rm R})$ is insignificant during the diffusion-dominated regime, the initial value of $J_a(0)$ along the universal path (which is the asymptotic value of the diffusion-dominated regime) can be determined by the above equation such that  $l^{\rm R}(J_a(0))= l_{i}^{\rm R} $ where we have reset time to $t=0$ for the evaluation of the universal path.
\begin{figure}[ht]
\begin{center}\includegraphics[width=0.79\textwidth]{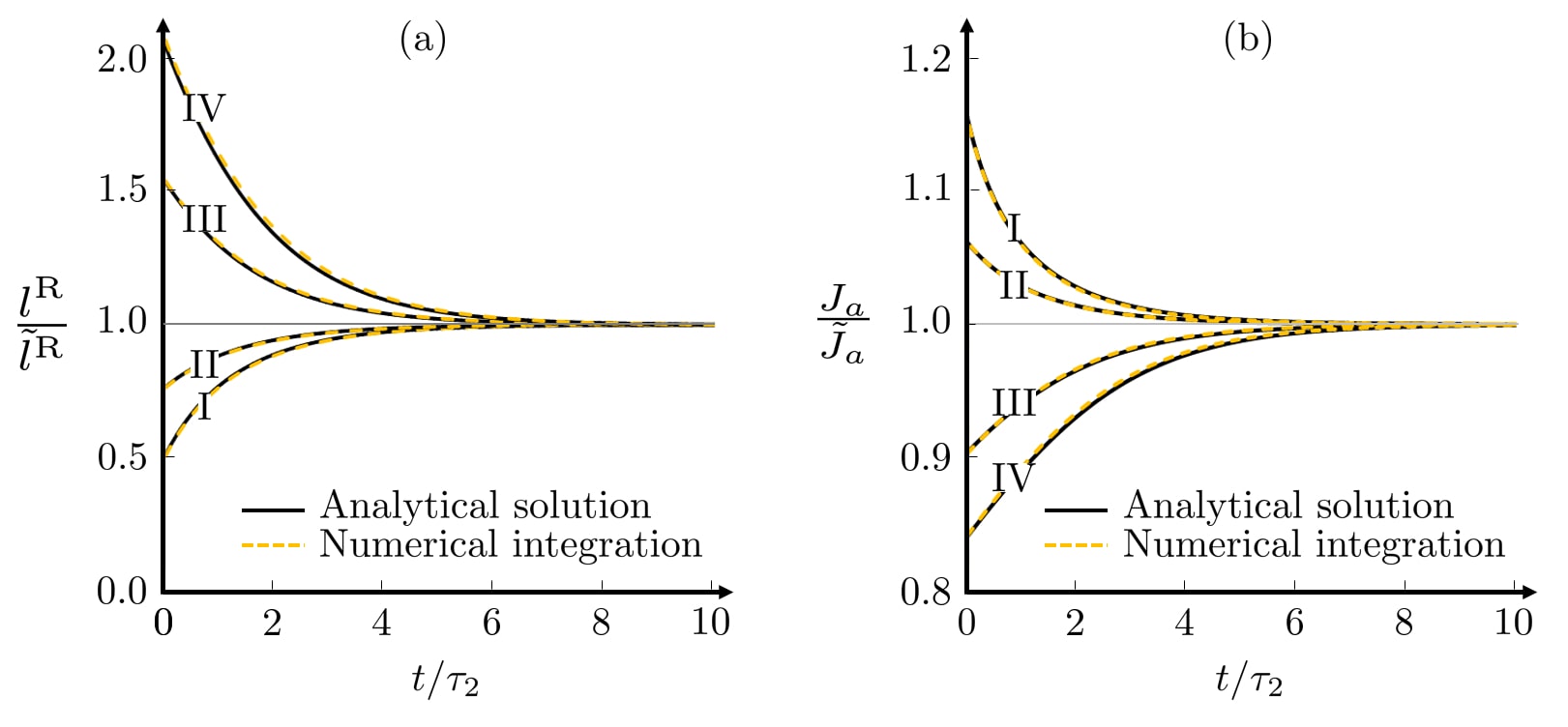}\end{center}
\caption{{Time evolution of (a) the normalized dry thickness and (b) the normalized swelling ratio at the association boundary along the universal path for initial conditions in regions I-IV presented in Fig. \ref{Phase}(a). Notice that here, $t=0$ corresponds to time at which evolution along the universal path begins. }}
\label{Tau2}
\end{figure}

\begin{figure}[ht]
\begin{center}\includegraphics[width=0.9\textwidth]{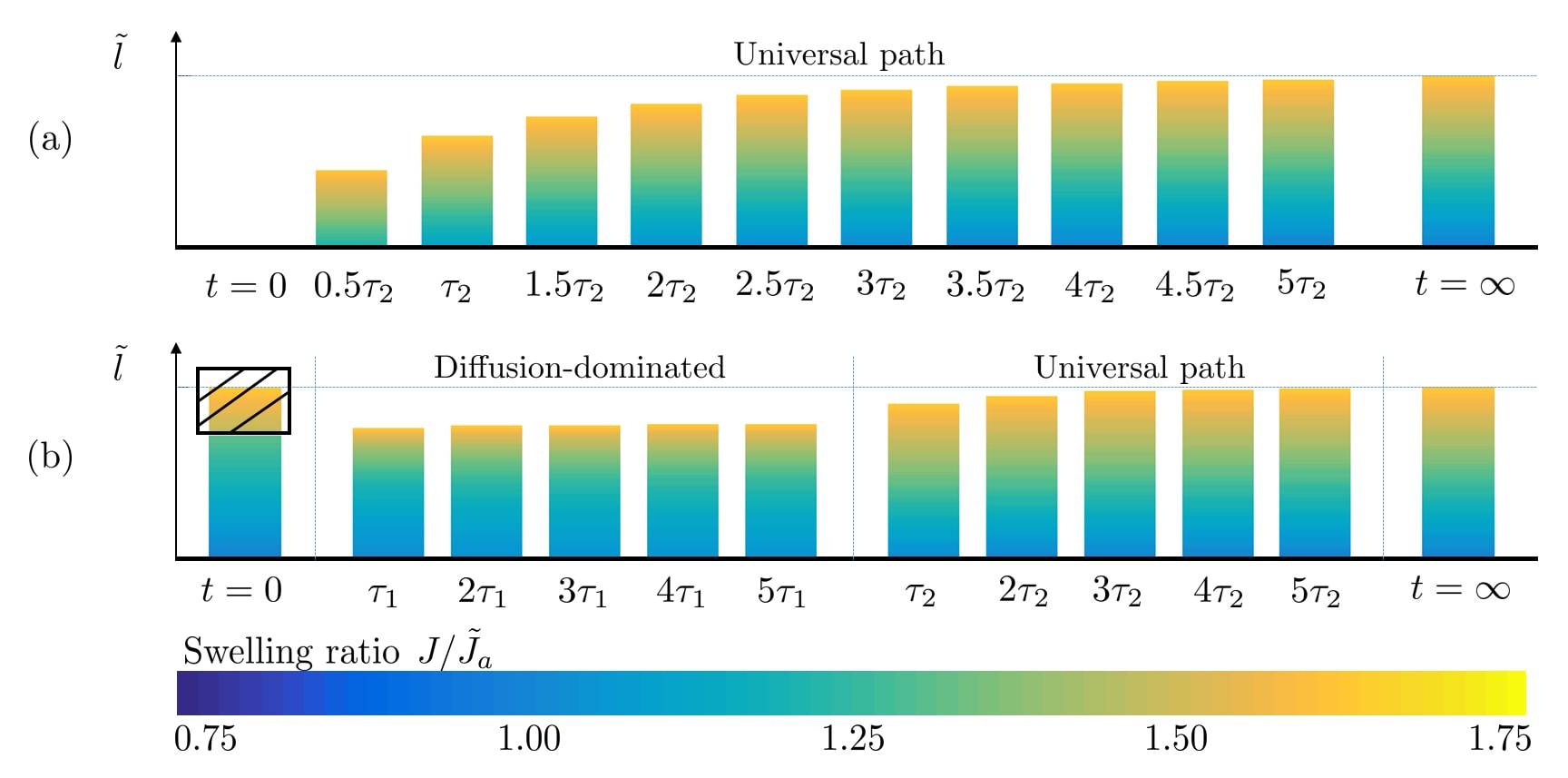} \end{center}
\caption{Time evolution of total thickness in the physical space and swelling ratio throughout the thickness for (a)  growth starting from inception  (b) a perturbed layer by the removal of a piece from the treadmilling regime, with $\psi_a/(kT/\nu)=-0.325$ and $\lambda_0=4.96$.}
\label{CutAPiece}
\end{figure}

The rate of change of the thickness can be determined by differentiating  the above equation. Then, comparing the differentiation result with    \eqref{lRVaVd} we arrive at a differential equation for $J_a=J_a(t)$, which by integration, reads
\begin{equation}\label{Jaintegral}
t=-\int_{J_a(0)}^{J_a(t)}\frac{F(J)}{{\rm V}_a^{\rm R}(J)+{\rm V}_d^{\rm R}}{d J }\quad \text{where} \quad F(J)=\left(\frac{l^{\rm R}(J)}{{\rm V}_a^{\rm R}(J)}\right)\frac{d{\rm V}_a^{\rm R}}{dJ}+ \frac{\lambda_0^4\Lambda(J)}{J{\rm V}_a^{\rm R}(J)}.
\end{equation}
\\
Once $J_a(t)$ is known, the variation of dry thickness is obtained by integration of equation \eqref{lRVaVd} to write
\begin{equation}\label{lRintegral}
{l^{\rm R}}(t) =l^{\rm R}_i+\int_{J_{u}}^{J_a(t)}\left({\rm V}_a^{\rm R}(J_a)+{\rm V}_d^{\rm R}\right){dJ_a}.
\end{equation}

In Fig. \ref{Tau2} we compare the time integration results of paths I - IV with the time-evolutions given by \eqref{Jaintegral} and \eqref{lRintegral}.  As shown in Fig. \ref{Tau2}, the integral analytical expressions are in good agreement with the numerical solutions in predicting the evolution of the swelling ratio and  dry thickness evolution.

\bigskip

\noindent{\bf Body inception and response to perturbation.} {In all the cases presented, growth has been studied with prescribed initial conditions. A specific challenge arises in defining initial conditions in absence of  solid material at the initiation of growth.  Inception of a new body, starting from the creation of the very first layer, ensues singularity, since at the time of  initiation the association  and dissociation surfaces overlap. Nonetheless, it is possible to take advantage of the universal path and to predict the growth of infinitesimally thin layers   in the limit $l^{\rm R}_i \rightarrow 0$. Employing
the analytical solutions  \eqref{Jaintegral} and \eqref{y-universal}  to predict the thickness of the body and the swelling ratio throughout the thickness at a given time,  we show on  Fig. \ref{CutAPiece}(a), the time evolution of a body from an initially nascent state. The evolution then follows the universal path}.

Moreover, our solutions suggest that if a perturbation occurs at any time during the evolution, the system will rapidly recover the universal path through a diffusion-dominated regime. Specifically, we show in Fig. \ref{CutAPiece}(b) the  response to  the perturbation of a treadmilling layer by  removal of a  portion of its thickness. It is observed that the perturbed body evolves back to treadmilling, after a transitory  diffusion-dominated regime  of time scale $\tau_1$ followed by evolution along the universal path of time scale $\tau_2$. Therefore, the treadmilling regime is the ultimate state towards which the system evolves naturally by  following the universal path.

\section{Conclusions}
\noindent A general theoretical framework for kinetics of surface growth coupled with solvent diffusion is presented. To account for conservation of mass in an open system, we consider the formation of a body that is composed of two species (i.e. an elastic matrix and a permeating solvent) and allow for association or dissociation of solvent units on its boundaries. An approach to determine the global natural reference configuration of the growing body in an arbitrary geometry is presented and utilized to determine the driving force of growth. Based on the dissipation inequality, restrictions on the kinetic law that relates between the growth rate and the driving force are  defined. As an illustration, the general framework is specialized to consider growth of polymer gels in a uniaxial setting. Numerical integration reveals that following a transient diffusion dominated response, the evolution of a body rapidly tends to a universal path, that is independent of initial conditions, and ultimately leads the system towards a treadmilling state. Any perturbation from the universal path is rapidly recovered, as swelling is adjusted throughout the body and the harmony between the two coupled mechanisms (i.e. growth and diffusion) is restored. The extension of our results to the limit in which the body is an infinitesimally thin layer yields an analytical prediction of the complete evolution of a body, from inception to treadmilling. There remains the philosophical question around the creation of the very first solid layer at the birth of the body, which is beyond the scope of the current framework. Finally, the universal path, that is derived here for a simple geometry, will be at the center of future investigations, which will facilitate this framework to explore more complex geometries. Experimental work will also be key to further illustrate this growth theory and to observe growth along the universal path in a laboratory setting.

\newpage
\section*{Appendix A - Nomenclature}\label{Nomenclature}
\noindent\textit{General framework:}
\begin{flushleft}\begin{tabular}{p{6mm} l }
$\mathcal{R}$ & body manifold in the physical configuration   \\
$\mathcal{R}^{\rm R}$ & material manifold in the reference configuration\\
$\partial \mathcal{R}$ & surface boundary of $\mathcal{R}$  \\
$\partial \mathcal{R}^{\rm R}$  & surface boundary of $\mathcal{R}^{\rm R}$\\
$\mathcal{S}_a$  & association surface in the physical configuration \\
$\mathcal{S}_a^{\rm R}$ & association surface in the reference configuration\\ $\mathcal{S}_d$ & dissociation surface in the physical configuration\\
$\mathcal{S}_d^{\rm R}$ & dissociation surface in the reference configuration\\
$t$ & time variable\\
${\bf y}$ & spatial point in the physical configuration\\
${\bf x}$ & material point in the reference configuration\\
$\hat {\bf y}$ &one on one mapping from reference configuration to physical configuration\\
$\hat {\bf x}$ & inverse one on one mapping from physical configuration to reference configuration\\
${\bf n}$ & outward pointing normal vector on the boundary $\partial \mathcal{R}$  in the physical configuration\\
${\bf n}^{\rm R}$ & outward pointing normal vector on the boundary\\
$\partial \mathcal{R}^{\rm R}$   & in the reference configuration\\
${dV_y}$ & volume element in the physical configuration\\
${dV_x}$ & volume element in the reference configuration\\
${dA_y}$ & area element in the physical configuration\\
${ dA_x}$ & area element in the reference configuration\\
${\bf v}$ & particle velocity in the physical configuration\\
${\bf V}$ & boundary velocity in the physical configuration\\
${\bf V}_{\rm G}$ & growth velocity in the physical configuration\\
${\bf V}^{\rm R}$ & boundary velocity in the reference configuration\\
${\bf j}$ & flux of solvent  in the physical configuration\\
${\bf j}^{\rm R}$ & flux of solvent  in the reference configuration\\
${\bf F}$ & deformation gradient\\
${J}$ & volume ratio\\
$\phi$ & physical volume fraction of solvent  in the body\\
$\phi^{\rm R}$ & referential volume fraction of solvent in the body\\
${\psi}$ & Helmholtz free energy of the body\\
${\psi_e}$ &Helmholtz elastic free energy associated with the solid matrix\\
${\psi_s}$ &Helmholtz free energy of the solvent\\
${\bf T}$ & Cauchy stress tensor\\
${\bf S}$ & Piola stress tensor\\
${\bf b}$ & body forces in the physical configuration\\
${\bf b}^{\rm R}$ & body forces in the reference configuration\\
${p}$ & hydrostatic pressure\\
${\mu}$ & chemical potential of the solvent\\
${f}$ & driving force on the boundary\\
${\Delta \psi }$ & latent energy of growth\\
${\mathcal G}$ & growth function \\
\end{tabular}\end{flushleft}
\noindent\textit{Specific problem:}
\begin{flushleft}\begin{tabular}{p{0.6cm} l  }
${k}$ & Boltzmann constant\\
${T}$ & temperature\\
${N}$ & number of polymer chains per unit volume\\
${\nu}$ & volume of a solvent unit\\
${\chi}$ & Flory-Huggins interaction parameter\\
${D}$ & diffusion coefficient\\

${b}$ & reaction constant\\
${\psi_0}$ & free energy of the unmixed solvent\\
${\psi_a}$ & potential  energy gain at the association boundary\\
${\lambda_0}$ &imposed in-plane stretch\\
\end{tabular}\end{flushleft}

\section*{Appendix B - Explicit Formula}
\newcounter{defcounter}
\setcounter{defcounter}{0}
\newenvironment{myequation}{%
\addtocounter{equation}{-1}
\refstepcounter{defcounter}
\renewcommand\theequation{B\thedefcounter}
\begin{equation}}
{\end{equation}}
\noindent In the following, we provide the full analytical formula that have been represented in condensed form in the main text.

The continuity of the chemical potential \eqref{potentialBC} is
\begin{myequation}\label{JdsteadyAppendix}
\ln \left(1-\frac{1}{J_d}\right)+\frac{1}{J_d}+\frac{\chi}{J_d^2}+{N\nu}\left(\frac{J_d}{\lambda_0^4}-\frac{1}{J_d}\right)=0.
\end{myequation}Equation \eqref{fJ} that relates the driving force to the swelling ratio can be further developed using \eqref{psi}-\eqref{psim}, \eqref{Dpsia}, \eqref{Dpsid} and \eqref{muJ} to write the driving force at the association boundary as
\begin{myequation}\label{fasteadyAppendix}
\begin{split}
f_{a}(J_{a})=\psi_0+\psi_a-\frac{kT}{\nu}\left(J_{a}-1\right)\left[\ln \left(1-\frac{1}{J_{a}}\right)+\frac{\chi}{ J_{a}}\right]-\frac{NkT}{2}\left(\frac{{J_{a}}^2}{\lambda_0^4}+2\lambda_0^2-3-2 \ln  J_{a}\right)\\+\left\{  \frac{kT}{\nu}\left[\ln \left(1-\frac{1}{J_{a}}\right)+\frac{1}{J_{a}}+\frac{\chi}{{J_{a}}^2}\right]+NkT\left(\frac{J_{a}}{\lambda_0^4}-\frac{1}{ J_{a}}\right)\right\} J_{a},
\end{split}
\end{myequation}
and at the dissociation boundary as
\begin{myequation}\label{fdsteadyAppendix}
f_d\left(J_{d}\right)=\psi_0 \left(1+J_d\right)-\frac{kT}{\nu}\left(J_d-1\right)\left[\ln \left(1-\frac{1}{J_d}\right)+\frac{\chi}{J_d}\right]-\frac{NkT}{2}\left(\frac{J_d^2}{\lambda_0^4}+2\lambda_0^2-3-2 \ln J_d\right).
\end{myequation}Substituting \eqref{VR} and \eqref{gfunction} in \eqref{y-universal} and \eqref{Deltax-universal}, we can write in the physical space
\begin{myequation}\label{Jy1steady}
\begin{split}
y_1({J})= \frac{\lambda_0^2D}{{b}\sinh\left(\frac{\nu}{kT}f_a(J_a)\right)}\bigg\{\frac{N\nu}{\lambda_0^4}\left[\ln \left(\frac{J}{J_{a}}\right)+\frac{1}{J}-\frac{1}{J_{a}}\right] -\frac{N\nu}{2}{\left(\frac{1}{ J^2}-\frac{1}{J_{a}^2}\right)}+\\
\frac{1}{3}\left( N\nu+2\chi-1\right)\left(\frac{1}{ J^3}-\frac{1}{ J_{a}^3}\right)-\frac{\chi}{2}\left(\frac{1}{ J^4}-\frac{1}{ J_{a}^4}\right)\bigg \},
\end{split}
\end{myequation}
and in the reference space
\begin{myequation}\label{Jx1steady}
\begin{split}
\Delta x_1({J})= \frac{\lambda_0^4D}{{b}\sinh\left(\frac{\nu}{kT}f_a(J_a)\right)} \bigg[\frac{N\nu}{\lambda_0^4} \left(-\frac{1}{J}+\frac{1}{J_a} +\frac{1}{2 J^2}-\frac{1}{2 {J_a}^2}\right) - \frac{N\nu}{3} \left(\frac{1}{J^3}-\frac{1}{ {J_a}^3}\right)+\\
\frac{1}{4}\left( N\nu+2\chi-1\right)\left(\frac{1}{ J^4}-\frac{1}{ J_{a}^4}\right)-\frac{2\chi}{5}\left(\frac{1}{ J^5}-\frac{1}{J_{a}^5}\right)\bigg].
\end{split}
\end{myequation}
Analytical expressions for \eqref{y-stationary} and \eqref{Deltax} can be obtained by replacing $J$ and $J_a$ in \eqref{Jy1steady} and \eqref{Jx1steady} by $\tilde J$ and $\tilde J_a$. Notice that the determination of $J_a$ is required to use the above relations: during the evolution along the universal path, the swelling ratio at the association boundary  $J_a(t)$ can be determined at any time using implicit equation \eqref{Jaintegral}; for the treadmilling regime, the swelling ratio at the association boundary $\tilde J_a$  is implicitly determined using the balance between addition and removal of mass \eqref{Jainf} combined combination with  the kinetic law \eqref{VR}, and satisfies
\begin{myequation}\label{Jasteady}
\begin{split}
f_{a}(\tilde J_{a})=-f_{d}(J_d).
\end{split}
\end{myequation}The thickness of the material, in both the physical and the reference frames, $l$ and $l^{\rm R}$, is obtained by taking the values of \eqref{Jy1steady} and \eqref{Jx1steady} respectively, for $J=J_d$.

\bibliographystyle{elsarticle-harv}
\bibliography{mybibfile}

\begin{thebibliography}{54}
\expandafter\ifx\csname natexlab\endcsname\relax\def\natexlab#1{#1}\fi
\expandafter\ifx\csname url\endcsname\relax
  \def\url#1{\texttt{#1}}\fi
\expandafter\ifx\csname urlprefix\endcsname\relax\def\urlprefix{URL }\fi

\bibitem[{Ambrosi et~al.(2011)Ambrosi, Ateshian, Arruda, Cowin, Dumais,
  Goriely, Holzapfel, Humphrey, Kemkemer, Kuhl, et~al.}]{Ambrosi2011}
Ambrosi, D., Ateshian, G., Arruda, E., Cowin, S., Dumais, J., Goriely, A.,
  Holzapfel, G.~A., Humphrey, J., Kemkemer, R., Kuhl, E., et~al., 2011.
  Perspectives on biological growth and remodeling. Journal of the Mechanics
  and Physics of Solids 59~(4), 863--883.

\bibitem[{Ambrosi and Mollica(2002)}]{Ambrosi2002}
Ambrosi, D., Mollica, F., 2002. On the mechanics of a growing tumor.
  International journal of engineering science 40~(12), 1297--1316.

\bibitem[{Ambrosi and Preziosi(2009)}]{Ambrosi2009}
Ambrosi, D., Preziosi, L., 2009. Cell adhesion mechanisms and stress relaxation
  in the mechanics of tumours. Biomechanics and modeling in mechanobiology
  8~(5), 397--413.

\bibitem[{Archer(2013)}]{archer2013}
Archer, R.~R., 2013. Growth stresses and strains in trees. Vol.~3. Springer
  Science \& Business Media.

\bibitem[{Ateshian(2007)}]{ateshian2007}
Ateshian, G.~A., 2007. On the theory of reactive mixtures for modeling
  biological growth. Biomechanics and modeling in mechanobiology 6~(6),
  423--445.

\bibitem[{Bau{\"e}r et~al.(2017)Bau{\"e}r, Tavacoli, Pujol, Planade, Heuvingh,
  and Du~Roure}]{bauer2017new}
Bau{\"e}r, P., Tavacoli, J., Pujol, T., Planade, J., Heuvingh, J., Du~Roure,
  O., 2017. A new method to measure mechanics and dynamic assembly of branched
  actin networks. Scientific reports 7, 15688.

\bibitem[{Bauhofer et~al.(2017)Bauhofer, Kr{\"o}del, Bilal, Daraio, and
  Constantinescu}]{bauhofer2017direct}
Bauhofer, A., Kr{\"o}del, S., Bilal, O., Daraio, C., Constantinescu, A., 2017.
  Direct laser writing of single-material sheets with programmable self-rolling
  capability. Bulletin of the American Physical Society 62.

\bibitem[{Ben~Amar and Ciarletta(2010)}]{amar2010}
Ben~Amar, M., Ciarletta, P., 2010. Swelling instability of surface-attached
  gels as a model of soft tissue growth under geometric constraints. Journal of
  the Mechanics and Physics of Solids 58~(7), 935--954.

\bibitem[{Ben~Amar and Goriely(2005)}]{amar2005}
Ben~Amar, M., Goriely, A., 2005. Growth and instability in elastic tissues.
  Journal of the Mechanics and Physics of Solids 53~(10), 2284--2319.

\bibitem[{Chester and Anand(2010)}]{Chester}
Chester, S.~A., Anand, L., 2010. A coupled theory of fluid permeation and large
  deformations for elastomeric materials. Journal of the Mechanics and Physics
  of Solids 58~(11), 1879--1906.

\bibitem[{Ciarletta et~al.(2013)Ciarletta, Preziosi, and Maugin}]{Ciarletta}
Ciarletta, P., Preziosi, L., Maugin, G., 2013. Mechanobiology of interfacial
  growth. Journal of the Mechanics and Physics of Solids 61~(3), 852--872.

\bibitem[{Correa et~al.(2015)Correa, Papadopoulou, Guberan, Jhaveri, Reichert,
  Menges, and Tibbits}]{correa}
Correa, D., Papadopoulou, A., Guberan, C., Jhaveri, N., Reichert, S., Menges,
  A., Tibbits, S., 2015. 3d-printed wood: Programming hygroscopic material
  transformations. 3D Printing and Additive Manufacturing 2~(3), 106--116.

\bibitem[{Cyron and Humphrey(2017)}]{Cyron}
Cyron, C., Humphrey, J., 2017. Growth and remodeling of load-bearing biological
  soft tissues. Meccanica 52~(3), 645--664.

\bibitem[{Dervaux and Ben~Amar(2011)}]{dervaux2011}
Dervaux, J., Ben~Amar, M., 2011. Buckling condensation in constrained growth.
  Journal of the Mechanics and Physics of Solids 59~(3), 538--560.

\bibitem[{DiCarlo(2005)}]{DiCarlo}
DiCarlo, A., 2005. Surface and bulk growth unified. In: Mechanics of material
  forces. Springer, pp. 53--64.

\bibitem[{Duda et~al.(2010)Duda, Souza, and Fried}]{Duda}
Duda, F.~P., Souza, A.~C., Fried, E., 2010. A theory for species migration in a
  finitely strained solid with application to polymer network swelling. Journal
  of the Mechanics and Physics of Solids 58~(4), 515--529.

\bibitem[{Feynman et~al.(1963)Feynman, Leighton, and Sands}]{Feynman}
Feynman, R.~P., Leighton, R.~B., Sands, M., 1963. The Feynman Lectures on
  Physics. Vol.~1. Elsevier.

\bibitem[{Flory(1942)}]{Flory42}
Flory, P.~J., 1942. Thermodynamics of high polymer solutions. The Journal of
  chemical physics 10~(1), 51--61.

\bibitem[{Flory and Rehner(1943)}]{Flory43}
Flory, P.~J., Rehner, J.~J., 1943. Statistical mechanics of cross-linked
  polymer networks ii. swelling. The Journal of Chemical Physics 11~(11),
  521--526.

\bibitem[{Fried and Gurtin(2004)}]{Fried04}
Fried, E., Gurtin, M.~E., 2004. A unified treatment of evolving interfaces
  accounting for small deformations and atomic transport with emphasis on
  grain-boundaries and epitaxy. Advances in applied mechanics 40, 1--177.

\bibitem[{Ganghoffer and Goda(2018)}]{ganghoffer2018}
Ganghoffer, J.-F., Goda, I., 2018. A combined accretion and surface growth
  model in the framework of irreversible thermodynamics. International Journal
  of Engineering Science 127, 53--79.

\bibitem[{Gibson et~al.(2014)Gibson, Rosen, and Stucker}]{gibson}
Gibson, I., Rosen, D., Stucker, B., 2014. Additive manufacturing technologies:
  3D printing, rapid prototyping, and direct digital manufacturing. Springer.

\bibitem[{Goriely(2017)}]{Goriely}
Goriely, A., 2017. The mathematics and mechanics of biological growth. Vol.~45.
  Springer.

\bibitem[{Gurtin(2008)}]{gurtin2008configurational}
Gurtin, M.~E., 2008. Configurational forces as basic concepts of continuum
  physics. Vol. 137. Springer Science \& Business Media.

\bibitem[{Holland et~al.(2013)Holland, Kosmata, Goriely, and Kuhl}]{Holland}
Holland, M.~A., Kosmata, T., Goriely, A., Kuhl, E., 2013. On the mechanics of
  thin films and growing surfaces. Mathematics and Mechanics of Solids 18~(6),
  561--575.

\bibitem[{Hong et~al.(2008)Hong, Zhao, Zhou, and Suo}]{Hong}
Hong, W., Zhao, X., Zhou, J., Suo, Z., 2008. A theory of coupled diffusion and
  large deformation in polymeric gels. Journal of the Mechanics and Physics of
  Solids 56~(5), 1779--1793.

\bibitem[{Humphrey et~al.(2014)Humphrey, Dufresne, and Schwartz}]{Humphrey}
Humphrey, J.~D., Dufresne, E.~R., Schwartz, M.~A., 2014. Mechanotransduction
  and extracellular matrix homeostasis. Nature reviews Molecular cell biology
  15~(12), 802--812.

\bibitem[{Kaur et~al.(1980)Kaur, Pandya, and Chopra}]{kaur}
Kaur, I., Pandya, D., Chopra, K., 1980. Growth kinetics and polymorphism of
  chemically deposited cds films. Journal of the Electrochemical Society
  127~(4), 943--948.

\bibitem[{Kim et~al.(2003)Kim, Jang, Kim, Cho, Yang, Kang, Min, and Lee}]{kim}
Kim, D.-H., Jang, H.-S., Kim, C.-D., Cho, D.-S., Yang, H.-S., Kang, H.-D., Min,
  B.-K., Lee, H.-R., 2003. Dynamic growth rate behavior of a carbon nanotube
  forest characterized by in situ optical growth monitoring. Nano Letters
  3~(6), 863--865.

\bibitem[{Loeffel and Anand(2011)}]{Loeffel}
Loeffel, K., Anand, L., 2011. A chemo-thermo-mechanically coupled theory for
  elastic--viscoplastic deformation, diffusion, and volumetric swelling due to
  a chemical reaction. International Journal of Plasticity 27~(9), 1409--1431.

\bibitem[{Louchev et~al.(2003)Louchev, Laude, Sato, and Kanda}]{louchev}
Louchev, O.~A., Laude, T., Sato, Y., Kanda, H., 2003. Diffusion-controlled
  kinetics of carbon nanotube forest growth by chemical vapor deposition. The
  Journal of chemical physics 118~(16), 7622--7634.

\bibitem[{Mane and Lokhande(2000)}]{mane}
Mane, R., Lokhande, C., 2000. Chemical deposition method for metal chalcogenide
  thin films. Materials Chemistry and Physics 65~(1), 1--31.

\bibitem[{Mattevi et~al.(2011)Mattevi, Kim, and Chhowalla}]{mattevi}
Mattevi, C., Kim, H., Chhowalla, M., 2011. A review of chemical vapour
  deposition of graphene on copper. Journal of Materials Chemistry 21~(10),
  3324--3334.

\bibitem[{Menzel and Kuhl(2012)}]{Menzel}
Menzel, A., Kuhl, E., 2012. Frontiers in growth and remodeling. Mechanics
  research communications 42, 1--14.

\bibitem[{Meyyappan et~al.(2003)Meyyappan, Delzeit, Cassell, and
  Hash}]{meyyappan}
Meyyappan, M., Delzeit, L., Cassell, A., Hash, D., 2003. Carbon nanotube growth
  by pecvd: a review. Plasma Sources Science and Technology 12~(2), 205.

\bibitem[{Mitchison and Cramer(1996)}]{mitchison}
Mitchison, T., Cramer, L., 1996. Actin-based cell motility and cell locomotion.
  Cell 84~(3), 371--379.

\bibitem[{Mogilner and Oster(1996)}]{Mogilner96}
Mogilner, A., Oster, G., 1996. Cell motility driven by actin polymerization.
  Biophysical journal 71~(6), 3030--3045.

\bibitem[{Mogilner and Oster(2003{\natexlab{a}})}]{Mogilner2003-Force}
Mogilner, A., Oster, G., 2003{\natexlab{a}}. Force generation by actin
  polymerization ii: the elastic ratchet and tethered filaments. Biophysical
  journal 84~(3), 1591--1605.

\bibitem[{Mogilner and Oster(2003{\natexlab{b}})}]{Mogilner2003-Motors}
Mogilner, A., Oster, G., 2003{\natexlab{b}}. Polymer motors: pushing out the
  front and pulling up the back. Current biology 13~(18), R721--R733.

\bibitem[{Moulton et~al.(2012)Moulton, Goriely, and Chirat}]{Moulton}
Moulton, D.~E., Goriely, A., Chirat, R., 2012. Mechanical growth and
  morphogenesis of seashells. Journal of theoretical biology 311, 69--79.

\bibitem[{Nair and Nair(1991)}]{nair}
Nair, M., Nair, P., 1991. Simplified chemical deposition technique for good
  quality sns thin films. Semiconductor science and technology 6~(2), 132.

\bibitem[{Noireaux et~al.(2000)Noireaux, Golsteyn, Friederich, Prost, Antony,
  Louvard, and Sykes}]{Noireaux}
Noireaux, V., Golsteyn, R., Friederich, E., Prost, J., Antony, C., Louvard, D.,
  Sykes, C., 2000. Growing an actin gel on spherical surfaces. Biophysical
  journal 78~(3), 1643--1654.

\bibitem[{Papastavrou et~al.(2013)Papastavrou, Steinmann, and
  Kuhl}]{Papastavrou}
Papastavrou, A., Steinmann, P., Kuhl, E., 2013. On the mechanics of continua
  with boundary energies and growing surfaces. Journal of the Mechanics and
  Physics of Solids 61~(6), 1446--1463.

\bibitem[{Skalak et~al.(1982)Skalak, Dasgupta, Moss, Otten, Dullemeijer, and
  Vilmann}]{Skalak82}
Skalak, R., Dasgupta, G., Moss, M., Otten, E., Dullemeijer, P., Vilmann, H.,
  1982. Analytical description of growth. Journal of Theoretical Biology
  94~(3), 555--577.

\bibitem[{Skalak et~al.(1997)Skalak, Farrow, and Hoger}]{Skalak97}
Skalak, R., Farrow, D., Hoger, A., 1997. Kinematics of surface growth. Journal
  of mathematical biology 35~(8), 869--907.

\bibitem[{Sozio and Yavari(2017)}]{sozio2017}
Sozio, F., Yavari, A., 2017. Nonlinear mechanics of surface growth for
  cylindrical and spherical elastic bodies. Journal of the Mechanics and
  Physics of Solids 98, 12--48.

\bibitem[{Szost et~al.(2016)Szost, Terzi, Martina, Boisselier, Prytuliak,
  Pirling, Hofmann, and Jarvis}]{szost}
Szost, B.~A., Terzi, S., Martina, F., Boisselier, D., Prytuliak, A., Pirling,
  T., Hofmann, M., Jarvis, D.~J., 2016. A comparative study of additive
  manufacturing techniques: Residual stress and microstructural analysis of
  clad and waam printed ti--6al--4v components. Materials \& Design 89,
  559--567.

\bibitem[{Tepole et~al.(2011)Tepole, Ploch, Wong, Gosain, and
  Kuhl}]{tepole2011}
Tepole, A.~B., Ploch, C.~J., Wong, J., Gosain, A.~K., Kuhl, E., 2011. Growing
  skin: a computational model for skin expansion in reconstructive surgery.
  Journal of the Mechanics and Physics of Solids 59~(10), 2177--2190.

\bibitem[{Theriot(2000)}]{Theriot}
Theriot, J.~A., 2000. The polymerization motor. Traffic 1~(1), 19--28.

\bibitem[{Thompson(1970)}]{Thompson}
Thompson, D.~W., 1970. On growth and form, 1917. Cambridge [Eng.]: University
  press. xv.

\bibitem[{Tomassetti et~al.(2016)Tomassetti, Cohen, and Abeyaratne}]{TCA}
Tomassetti, G., Cohen, T., Abeyaratne, R., 2016. Steady accretion of an elastic
  body on a hard spherical surface and the notion of a four-dimensional
  reference space. Journal of the Mechanics and Physics of Solids 96, 333--352.

\bibitem[{Treloar(1975)}]{Treloar}
Treloar, L. R.~G., 1975. The physics of rubber elasticity. Oxford University
  Press, USA.

\bibitem[{Z{\"o}llner et~al.(2012)Z{\"o}llner, Tepole, and Kuhl}]{zollner2012}
Z{\"o}llner, A.~M., Tepole, A.~B., Kuhl, E., 2012. On the biomechanics and
  mechanobiology of growing skin. Journal of theoretical biology 297, 166--175.

\bibitem[{Zurlo and Truskinovsky(2017)}]{zurlo2017printing}
Zurlo, G., Truskinovsky, L., 2017. Printing non-euclidean solids. arXiv
  preprint arXiv:1703.03082.

\end{thebibliography}

\end{document}